\documentclass[useAMS,usenatbib]{mn2e}
\usepackage{textcomp}
\usepackage{amstext}
\usepackage{graphicx}

\def\sun{\hbox{$\odot$}}
\def\lesssim{\mathrel{\hbox{\rlap{\hbox{\lower4pt\hbox{$\sim$}}}\hbox{$<$}}}}
\def\gtrsim{\mathrel{\hbox{\rlap{\hbox{\lower4pt\hbox{$\sim$}}}\hbox{$>$}}}}

\newcommand{\mamo}[1]{\mbox{$#1$}}
\newcommand{\unit}[1]{\ifmmode \:\mbox{\rm #1}\else \mbox{#1}\fi}
\newcommand{\sbr}[1]{_{\rm #1}}
\newcommand{\mone}{\mamo{^{-1}}}

\newcommand{\kms}{\unit{km~s\mone}}
\newcommand{\kpc}{\unit{kpc}}
\newcommand{\mpc}{\unit{Mpc}}
\newcommand{\hkpc}{\mamo{h\mone}\kpc}
\newcommand{\hmpc}{\mamo{h\mone}\mpc}

\newcommand{\secref}[1]{Section~\ref{sec:#1}}

\newcommand{\figref}[1]{Fig.~\ref{fig:#1}}
\newcommand{\tabref}[1]{Table~\ref{tab:#1}}

\begin{document}

\title{Quenching star formation in cluster galaxies}

\author[Dan S. Taranu et al.]{Dan S. Taranu $^{1,2}$, Michael J. Hudson $^{2,3}$, Michael L. Balogh $^{2,4}$, Russell J. Smith $^{5}$, \newauthor Chris Power $^{6,7}$, Kyle A. Oman $^{8,2}$, Brad Krane $^{2}$\\
$^{1}$Department of Astronomy \& Astrophysics, University of Toronto, 50 St. George Street, Toronto, ON, Canada, M5S 3H4.\\
$^{2}$Department of Physics and Astronomy, University of Waterloo, 200 University Avenue West, Waterloo, ON, Canada, N2L 3G1.\\
$^{3}$Perimeter Institute for Theoretical Physics, 31 Caroline St. N., Waterloo, ON, N2L 2Y5, Canada.\\
$^{4}$Leiden Observatory, Leiden University, PO Box 9513, 2300 RA Leiden, The Netherlands.\\
$^{5}$Department of Physics, University of Durham, Science Laboratories, South Road, Durham DH1 3LE.\\
$^{6}$International Centre for Radio Astronomy Research, The University of Western Australia, 35 Stirling Highway, Crawley, WA 6009, Australia.\\
$^{7}$ARC Centre of Excellence for All-Sky Astrophysics (CAASTRO).\\
$^{8}$Department of Physics and Astronomy, University of Victoria, 3800 Finnerty Rd, Victoria, BC, V8P 5C2.\\
}

\maketitle
\begin{abstract}

In order to understand the processes that quench star formation within rich clusters,  we construct a library of subhalo orbits drawn from $\Lambda$CDM cosmological N-body simulations of four rich clusters. The orbits are combined with models of star formation followed by quenching in the cluster environment. These are compared with observed bulge and disc colours and stellar absorption linestrength indices of satellite galaxies. 
Models in which the bulge stellar populations depend only on the galaxy subhalo mass while the disc quenching depends on the cluster environment are acceptable fits to the data. An exponential disc quenching timescale of 3 -- 3.5 Gyr is preferred. Models with short ($\lesssim 1$ Gyr) quenching timescales yield cluster-centric gradients in disc colours and Balmer line indices that are too steep compared to observations. We also examine models in which there is quenching in lower mass groups prior to cluster infall (``pre-processing''), finding that such models are a better fit to the data than models without pre-processing and require similar quenching times. The data slightly prefer models where quenching occurs only for galaxies falling within about 0.5 $r_{200}$. Finally, we have examined models with short quenching timescales of 1 Gyr, but a long delay time of 3 Gyr prior to quenching. All models with short quenching timescales, even such ``delayed-then-rapid'' quenching models, produce excessively red galaxies near the cluster core and are strongly disfavoured by the data. These results imply that the environments of rich clusters must impact star formation rates of infalling galaxies on relatively long timescales -- several times longer than a typical halo spends within the virial radius of a cluster. This scenario favours gentler quenching mechanisms such as slow ``strangulation'' over more rapid ram-pressure stripping.
\end{abstract}


\section{Introduction}

Understanding the physical mechanisms that halt star formation in galaxies has been a long-standing challenge in galaxy formation and evolution.  It is clear from the colour-magnitude relation \citep{SanVis78, BowLucEll92} that mass or, more accurately, velocity dispersion \citep{SmiLucHud09c, GraFabSch09} is a driving parameter of the stellar populations in red galaxies.  At the same time, the morphology--density \citep{HubHum31, Dre80, PosGel84}, the colour--density \citep{BalBalNic04, HogBlaBri04} and the age-density \citep{SmiHudLuc06} relations are strong evidence of the role of environment.  The latter suggest that environment somehow contributes to ``quenching'' star formation, moving galaxies from the blue cloud to the red sequence. Quenching can be abrupt or even instantaneous, as in dousing a fire, or it can operate on longer timescales. It has become common to model both an ``internal'' quenching mechanism (assumed to be tied to galaxy mass) as well as environmental quenching, associated with a galaxy falling into a higher mass halo and becoming a satellite \citep{KauWhiGui93, ColLacBau00, vanAquYan08, WeiKauvan09, PenLilKov10, PenLilRen11,WetTinCon13}. 

Many mechanisms that might quench star formation in cluster environments have been proposed: ram-pressure stripping of the cold gas \citep{GunGot72, AbaMooBow99}, the removal of the hot gas halo \citep{LarTinCal80}, sometimes known as ``strangulation'' \citep{BalMor00} either via ram-pressure \citep{McCFreFon08} or via tidal stripping by the cluster potential \citep{Mer84, Mam87}, or by ``harassment'' -- encounters with other galaxies \citep{GalOst72, MooLakKat98}. The ram-pressure stripping and strangulation mechanisms remove the gas that is the fuel for star formation, and so the stellar disc will dim and redden as its stars age.  Several of the processes affecting late-type galaxies have been reviewed by \cite{BosGav06}.  

It has long been known that spiral galaxies in the cores of rich clusters are deficient in HI \citep{DavLew73, 
HayGioChi84, GioHay85, SolManGar01} and stripping of star-forming gas has been observed in nearby rich clusters \citep{KooKen04, ChuvanGKen07, SunDonVoi07, YagKomYos07, YosYagKom08, SmiLucHam10, SivRieRie10}. This stripping usually appears to occur on short (several hundred Myr) timescales, but it is unclear if this is the dominant quenching mechanism. Other, less abrupt or ``gentler'' mechanisms such as strangulation or tidal stripping are more difficult to observe directly. To determine which mechanisms are dominant, one might instead tackle the following more empirical questions:  At what location in the cluster or group does quenching first occur?  What is the timescale of quenching - is quenching abrupt, ending all star formation within tens of millions of years, or does star formation slowly decline over billions of years? What range of satellite galaxy masses are effectively quenched by the environment? Are all morphological components quenched equally, or, for example, are bulges insensitive to environment while discs are quenched?  

A common approach to understanding these processes is semi-analytic modelling (SAMs for short; see \cite{Bau06} for a review). SAMs attempt to model important physical process via analytical prescriptions, combining these with results from numerical experiments like N-body simulations (hence the semi). A disadvantage of the SAM approach when applied to all galaxy populations is that it is difficult to isolate and study specific environments and physical processes, particularly with a large number of potentially correlated free parameters and physical processes. It is also challenging to determine if the parametrisations of physical processes are themselves appropriate. An alternative approach is to apply only a minimal number of physically-motivated prescriptions relevant to a specific environment, rather than all possible environments. We choose to focus on the cluster environment. \cite{HudSteSmi10} have shown that cluster galaxies have bulge fractions and bulge properties which are largely independent of environment and dependent only on internal structure of the galaxy, and that environmental trends are driven by variations in properties of the disc. This forms the basis of our simple model for quenching, outlined below.

In this paper, we will examine the link between star formation rates and the cluster environment by combining n-body simulations of rich clusters with star formation models. Previous models of galaxy evolution in a clusters \citep[and references therein]{BalNavMor00, DiaKauBal01} employed models linking star formation rates with gas consumption, and in some cases, replenishment \citep{WeiKauvon09}. More recent studies \citep{BerSteBul09, McGBalBow09, DeWeiPog11, SmiLucPri12} have focused on pre-processing in galaxy groups. While these approaches have been successful in reproducing star formation rates (SFR) and colour gradients, they have not yet isolated the mechanism responsible for halting star formation despite including a wide variety of models for physical processes. We instead seek to produce models that employ an easily-tested causal connection between environment (traced by infall into a rich cluster) and the evolution of an individual galaxy (see \citealt{MahMamRay11} for a recent example of star formation models of cluster galaxies). 

In our models, galaxies (or components thereof such as the bulge and disc) are assumed to be forming stars until quenched by one of two mechanisms. In one scenario, ``internal'' quenching of star formation occurs at a time which is assumed to be a function of a galaxy's internal properties (specifically velocity dispersion), such that galaxies with high velocity dispersions are quenched earlier (see \secref{ahsmodels} for details of our particular implementation). In the alternative scenario, environmental quenching of star formation occurs in the galaxy when its halo crosses some fraction of the virial radius of the cluster, with an option to delay quenching until the galaxy reaches pericentre or for some fixed time after infall. Though these prescriptions depict a simplified view of galaxy evolution, testing models based upon these mechanisms should provide clues as to the relative importance of environments in star formation. Varying the quenching radius of the cluster and allowing quenching to occur in smaller, group-sized halos (``pre-processing'') will also help to determine where quenching is likely to occur.

Most models to date have compared predictions with observations of galaxy luminosities and colours. We aim to compare our models not just with total galaxy colours - which are very red for most cluster galaxies - but also with disc colours, which are more sensitive to star formation models and thus better discriminants of quenching models. We also provide predictions of spectroscopic absorption linestrength indices, which help to break the well-known degeneracies between age, metallicity and $\alpha$-element enhancement. In this paper, we focus on colour and absorption linestrength data from the NOAO Fundamental Plane Survey \citep[hereafter NFPS]{SmiHudNel04} of galaxies in rich clusters, as well as bulge-disc decompositions with more extended coverage beyond $r_{200}$ from SDSS \citep{SimMenPat11}.

This paper is structured as follows. \secref{nbody} details the simulations used and presents information on the simulated clusters. \secref{data} describes the NFPS and SDSS data in more detail.  \secref{sfmodels} describes the star formation models employed and compares the resulting trends in galaxy colours with observations. 
The predicted linestrengths are compared to observations in \secref{lineindices}. We discuss the implications of our results in \secref{discuss} and summarise and conclude in \secref{conclusions}. Further discussions of merger prescriptions and simple star formation models are available in Appendix \ref{sec:simplemergers}.

\section{N-Body simulations and halo orbits}
\label{sec:nbody}

\subsection{Methods}
\label{sec:methods}
The simulation data consists of four rich clusters generated using the publicly available parallel Tree-PM code GADGET-2 \citep{Spr05}. Cluster candidates were identified from a low-resolution initial simulation using $256^{3}$ particles in a cube of side length $512 \hmpc$. $\Lambda$CDM cosmological parameters $\Omega_{\Lambda}=0.72$, $\Omega_{m}=0.28$, $h=0.72$ and $\sigma_{8}=0.8$ were adopted, consistent with WMAP 7-year results \citep{KomSmiDun11}. Four candidate clusters were chosen for a higher-resolution ``zoom'' re-simulation \citep{KatWhi93,NavWhi94,PowNavJen03}, using 150 approximately equally-spaced time steps from $z=3$ to $z=0$. A fifth candidate was excluded due to a complex, late-time merger, which is atypical of cluster growth histories.  In brief, only particles passing near the central overdensity in each simulation are fully sampled. Particles which remain sufficiently far (in practice, more than about 20 Mpc/h) are subsampled and given higher masses to provide an accurate tidal field, with four levels of refinement up to the full $512 \hmpc$ side box. Each re-simulation has 12.5 million dark particles, 8.8 million of which are full resolution with a mass of $6.16 \times 10^8 M_{\sun}/h$. The particle mass and minimum gravitational softening length of $1 \hkpc$ are sufficient to resolve a Milky Way or M31 mass halo with at least 1,000 particles and achieve a resolution limit of 30 particles per halo for halos close to the total mass of a dwarf galaxy like the Large Magellanic Cloud.

The masses and other physical properties of each cluster at $z=0$ are given in \tabref{clusters}. Subhalo catalogues at each time step were generated using the AMIGA Halo Finder \citep[hereafter AHF]{KnoKne09}. AHF calculates isodensity contours on an adaptive mesh grid and identifies subhalos as collections of mutually bound particles within unambiguous density contours (i.e. overdensities). The resulting subhalo statistics are given in \tabref{clusters}. AHF determines halo sizes from upturns in radial density contours, using an iterative process to remove unbound particles. We create halo merger trees by linking between consecutive snapshots' subhalo catalogues.

A typical problem with N-body simulations of dissipationless dark matter is the overmerging of low-mass dark matter subhalos as they are tidally disrupted by the cluster. This effect has been shown to be at the least strongly resolution-dependent \citep{MooKatLak96b}, and mostly likely an artefact of poor resolution \citep{KlyGotKra99}. A common solution to this ``overmerging'' problem, adopted here, is to use the most bound particle(s) to track the orbits of these `orphan' haloes \citep{KauColDia99,SprWhiTor01}. In our implementation, any halo which is detected in at least one snapshot is tracked through its most bound particle in all subsequent snapshots, even if it is not detected again by AHF. Thus once a halo is detected at high redshift, it is guaranteed to still be present in the final $z=0$ snapshot. Although AHF is capable of detecting most of the massive subhalos in the simulation, the most bound particle tracking technique recovers on average four times as many halos as in the original halo AHF catalogue. 

Although this method recovers many genuinely independent subhalos, it can also recover subhalos that have genuinely merged. Merging is generally suppressed in massive clusters, since the dynamical friction timescale is long enough that satellites can survive for billions of years \citep{TorDiaSye98}, unlike in lower mass clusters \citep{Tor97}. However, merging may not be suppressed in the field prior to infall and may even be enhanced in groups; thus, some of the ``overmerged'' halos that were ``un-merged'' by this method may indeed have been genuine mergers prior to cluster infall. Semi-analytical models typically employ prescriptions for dynamical friction and merging to identify such unresolved potential mergers. Our simulations can adequately resolve dynamical friction but are limited in identifying substructure in large overdensities. Potential ``merger candidates'' are identified as those orphan subhalos at less than 75 kpc distance from a more massive parent halo and with a low binding ratio ($v/v\sbr{escape}<$0.1) at the final time step. While these cuts are somewhat arbitrary and are likely an overestimate of the true number of merged halos, since some satellites could be at similar distances without having yet been merged, the final consequence of including or excluding these candidate mergers on our results is generally small. We opt to exclude ``candidate merger'' subhalos from our analysis by default. A more detailed comparison of simplistic models with and without this merging prescription is presented in Appendix \ref{sec:simplemergers}.

Besides tracking overmerged halos, we use the most bound particles to fill in ``gaps'' in each halo's orbital history, e.g. for halos detected in the 8th and 10th snapshot but not the 9th, which can occur if halos briefly cross dense regions at pericenter, for example.

In this paper, we will typically define ``infall'' to occur when the subhalo crosses the virial radius $r_{200}$, defined to be the radius within which the mean density is 200 times the critical density, or in some models at a smaller fraction of $r_{200}$. We track $r_{200}$ across all redshifts, so the value evolves and is not fixed to $r_{200}$ at z=0. The mean cluster density thus drops by a factor of about 3 from z=1 to the present. The lookback time at the first $r_{200}$-crossing is noted as $t_{r_{200}}$.  The orbits allow us to also define the lookback time to first pericentre passage. For the rich clusters studied here, we find that, for recent infalls, the typically time between infall and pericentre is close to 1 Gyr.

In principle, pericentre can occur anywhere within $r_{200}$; in practice, few halos pass closer than $0.1 r_{200}$. The median pericentre is $0.4 r_{200}$ with no dependence on $t\sbr{infall}$ except at $z \la 0.1$, 
where it begins to dip to $0.3 r_{200}$.

\begin{table}
\begin{minipage}{9cm}
\begin{tabular}{|ccccccc|}
\hline 

$M\sbr{vir}$ & $R_{200}$ & $\sigma_{1D}$ & $N\sbr{P}$ \footnote{Millions of particles within the virial radius of the cluster} & $N\sbr{H}$ \footnote{Number of halos ever tracked, including recovered `orphan' halos.} & $N\sbr{I}$ \footnote{Number of halos which have ever crossed the virial radius of the cluster.} & $N\sbr{B}$ \footnote{Number of halos which crossed the virial radius but are now outside of it.} \\ 
$10^{14}M\sun/h$ & Mpc/$h$ & \kms & $10^6$ & & & \\
\hline
$17.90$ & 1.98 & 1176 & 2.9 & 33567 & 9328 & 1000\\ 
$13.89$ & 1.82 & 1103 & 2.3 & 32605 & 6468 & 1437\\ 
$9.11$ & 1.58 & 1073 & 2.7 & 30940 & 7511 & 1071\\ 
$9.82$ & 1.62 & 945 & 1.6 & 32980 & 4800 & 989\\ 
\hline
\end{tabular}
\end{minipage}
\caption{Properties of the four simulated rich clusters and their subhalo populations at $z=0$.
\label{tab:clusters}
}
\end{table}
In addition to tracking subhalo positions and velocities after infall, bound particles are used to compute velocity dispersions for each subhalo. Subhalos are assigned a \emph{maximum} velocity dispersion $\sigma\sbr{max}$ corresponding to the velocity dispersion at the time when the subhalo had the largest mass (and in turn had the largest amount of bound particles) prior to infall within $r_{200}$. This is based on the assumption that a subhalo will not grow significantly after infall, which is expected to be true. In the real universe, we might expect the galaxy's baryonic component, which has dissipated and become more tightly-bound, to be much less sensitive to tidal disruption than a dissipationless dark matter halo. Thus the maximum pre-infall halo velocity dispersion serves as a proxy for the (post-infall) stellar velocity dispersion \citep{ConWecKra06}, which in turn is the primary driver of the stellar populations (see \secref{sfmodels} below).

The simulations allow us to mimic observational samples by viewing the subhalos positions and velocities in projection. In all plots of projected quantities, we use the three axial projections of each cluster and appropriate line-of-sight velocities and velocity dispersions to average over projection effects and improve galaxy statistics. We denote the projected radius on the sky as $R$, to distinguish it from the 3D radius $r$. We also apply cuts in velocity-space, including all subhalos with velocity $cz$ within $\pm 3 \sigma\sbr{cl}$ of the cluster's mean $\overline{cz}$. Thus in the plots shown in projection, about 17 percent of the plotted subhalos have $R < r_{200}$ although in 3D they are outside the virial radius ($r > r_{200}$). This fraction can vary from 12-20 percent with velocity dispersion or infall-time related cuts, which we will apply when populating halos with galaxies in \secref{sfmodels}. For a more detailed discussion of contamination due to projection, see \cite{MamBivMur10}.

\subsection{Results}
\label{sec:nbodyresults}

We now examine cluster-centric trends of the two key subhalo properties to be passed to our galaxy evolution code in \secref{sfmodels}: subhalo velocity dispersion $\sigma\sbr{max}$  and infall/pericentre time. In our models, these two quantities will determine the star formation history of the galaxy stellar populations. 

\begin{figure*} 
\includegraphics[width=0.49\textwidth]{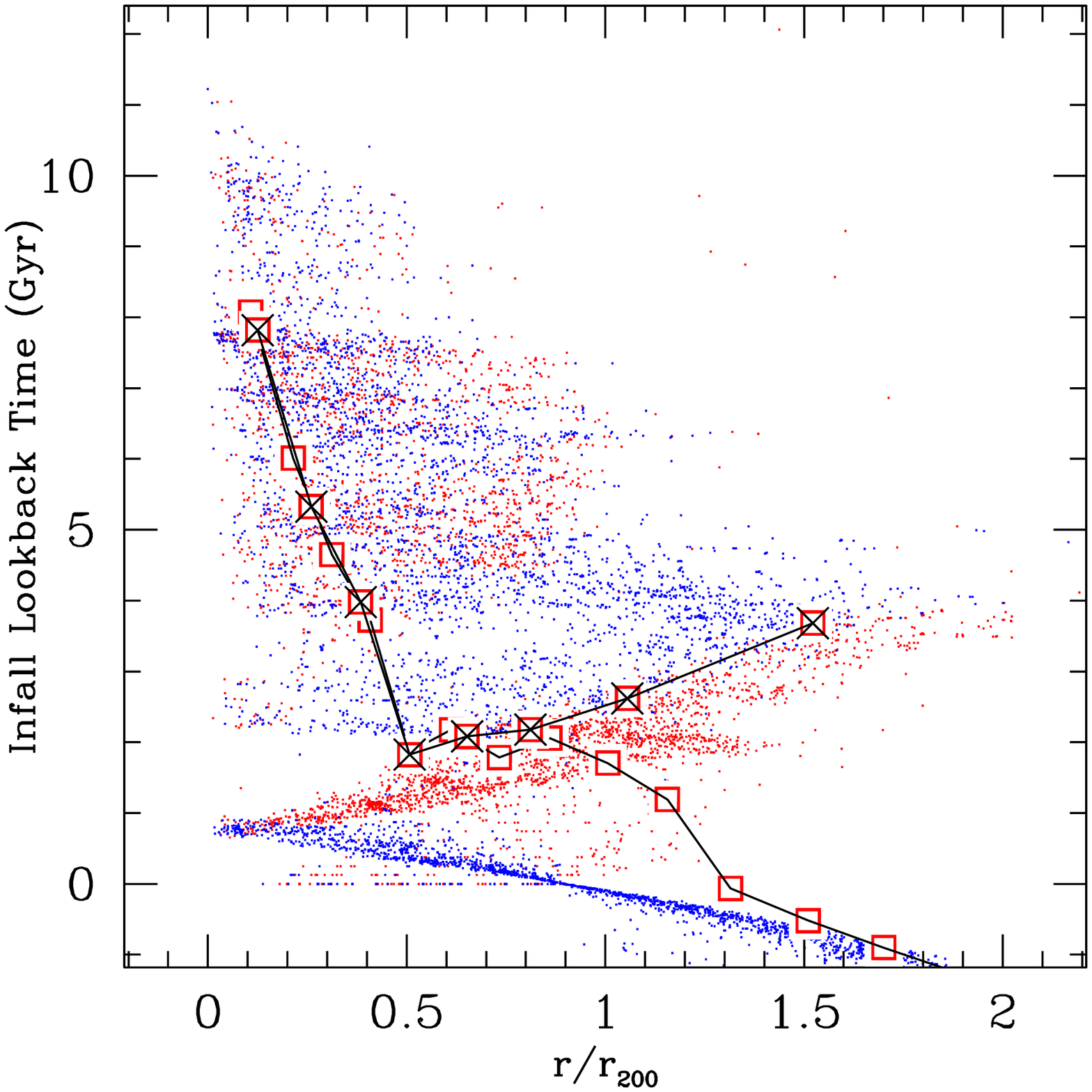}
\includegraphics[width=0.49\textwidth]{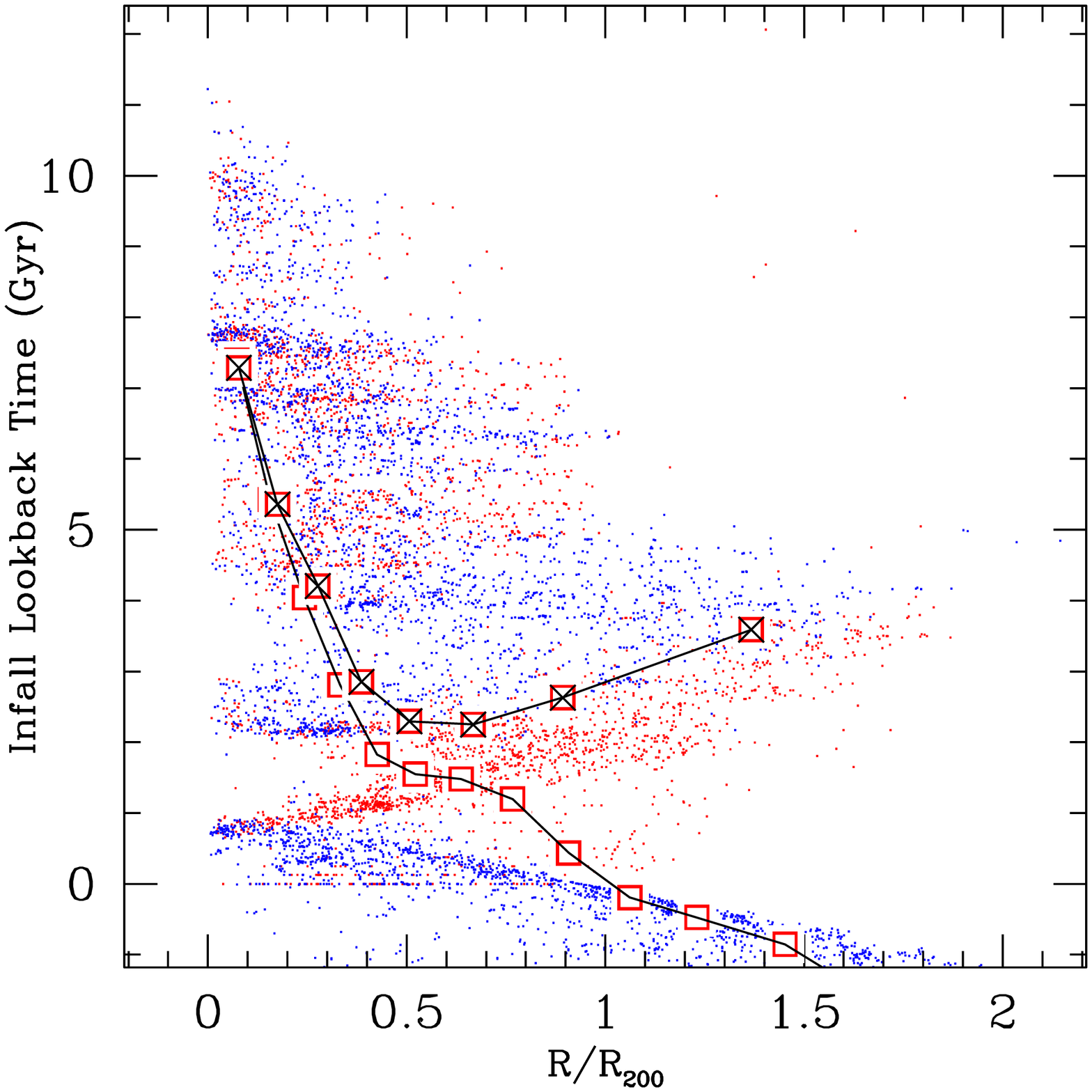}
\caption{Lookback time at infall ($t\sbr{r200}$) for subhalos as a function of radius, in three dimensions (left panel) and in projection (right panel). Halos are colour coded on whether they are approaching (blue) or receding from (red) the cluster. Halos that have not fallen in yet are given a naive estimate of their eventual infall time assuming no changes in their current radial velocity and the size of the cluster. Empty squares show the median infall times for all halos; squares with an x inside show median infall times for only those halos that have fallen in ($t\sbr{r200} > 0$).
\label{fig:tinfall_position}
} 
\end{figure*}

\figref{tinfall_position} shows the distribution of subhalo infall times past $r_{200}$ (i.e., time since first crossing of $r_{200}$) as a function of both projected and real distance to the centre of each cluster. We define accreted halos as those that have crossed $r_{200}$ of the cluster at least once. Most recently accreted halos ($<2$ Gyr) are clustered around the same position for any given infall time, likely because most halos have mostly radial orbits and thus have similar infall histories. The majority of halos with intermediate infall times (2-4 Gyr) are at large radii, even outside the virial radius of the cluster and thus form a ``backsplash'' population of galaxies that fell into and then back out of the cluster. Few backsplash galaxies remain so longer than 6 Gyr; such old accreted halos are concentrated near the centre of the cluster. This virialised population of subhalos outnumbers the very recent infalls within $0.25r_{200}$. At $0.5r_{200}$, the median infall time of a subhalo is barely 2 Gyr.

When considering only accreted halos, the backsplash population dominates at large radii ($>0.5r_{200}$ and the median infall time increases once again. However, when considering all halos and not just accreted halos, the median infall time continues to decrease past $r_{200}$, simply because the population outside $r_{200}$ is dominated by halos about to fall in for the first time, rather than halos exiting the cluster after reaching pericenter. These soon-to-be-accreted halos can be assigned zero or negative infall times based on estimates for their inevitable infall time onto the cluster (\figref{tinfall_position}), where a negative median infall time in a particular bin indicates that more than half of the halos have yet to fall in. The tight linear relation between these estimated infall times and $r/r_{200}$ suggests that most field galaxies within $r_{200}$ are falling in to the cluster on nearly direct radial orbits. As a result and despite the scatter in pericentres, it is clear in 3D that the accreted halo population can be separated by radial velocity up to an infall time of 2 -- 3 Gyr, where these recently accreted halos begin to reach apocentre. Halos with $t_{r_{200}} > 4$ Gyr are almost all part of the virialised population, since few are backsplash galaxies and there is a roughly even split between approaching and receding galaxies. 

The trends identified in 3D are, of course, less evident in projection. However, the clear distinction between approaching and receding halos remains. Similarly, there are two triangularly-shaped 'holes' in the projected phase space, with virtually no halos having infall times between 1 -- 2 Gyr within $0.5R_{200}$ and very few backsplash halos with $t\sbr{r200}< 2$ Gyr. If there is a strong environmental impact on infalling galaxies, there should be two distinct populations beyond $R_{200}$ - currently infalling halos hosting field galaxies and a smaller backsplash population hosting quenched galaxies. Within $R_{200}$, the distinction is less clear - halos have a variety of infall times, and some are simply within $R_{200}$ in projection but have yet to cross $r_{200}$. These projection effects serve to lower the median infall time. As we will demonstrate in \secref{singlemodels}, the trends identified in infall time can be mapped onto a stellar population age for halos hosting galaxies undergoing infall-based quenching, which is particularly appropriate for discs. In particular, the sharp drop in median infall time between 0.8-1 $R_{200}$ will cause a similarly large change in optical colours of galaxies for models employing abrupt quenching on infall.

These trends are also mass-dependent, as will be demonstrated shortly. At any given radius, the median infall time is larger for massive halos, and so the trends shown in \figref{tinfall_position} become smoother when smaller halos (which are unlikely to host massive galaxies) are excluded. The small population of halos close to the cluster center but with very small infall times ($r<r_{200}$, $t_{infall}<$0.3 Gyr) are mostly spurious numerical artefacts, which tend to be very low mass halos at the resolution limit and not real galaxies. Finally, it should be noted that halos can also fall in to a group before reaching the cluster. Groups of multiple halos can then fall in to the cluster nearly simultaneously, and a few such groups are visible as bands of halos at very similar infall times in \figref{tinfall_position}. This effect is partly diluted by the superposition of data from four different clusters.

\begin{figure*}
\includegraphics[width=\textwidth]{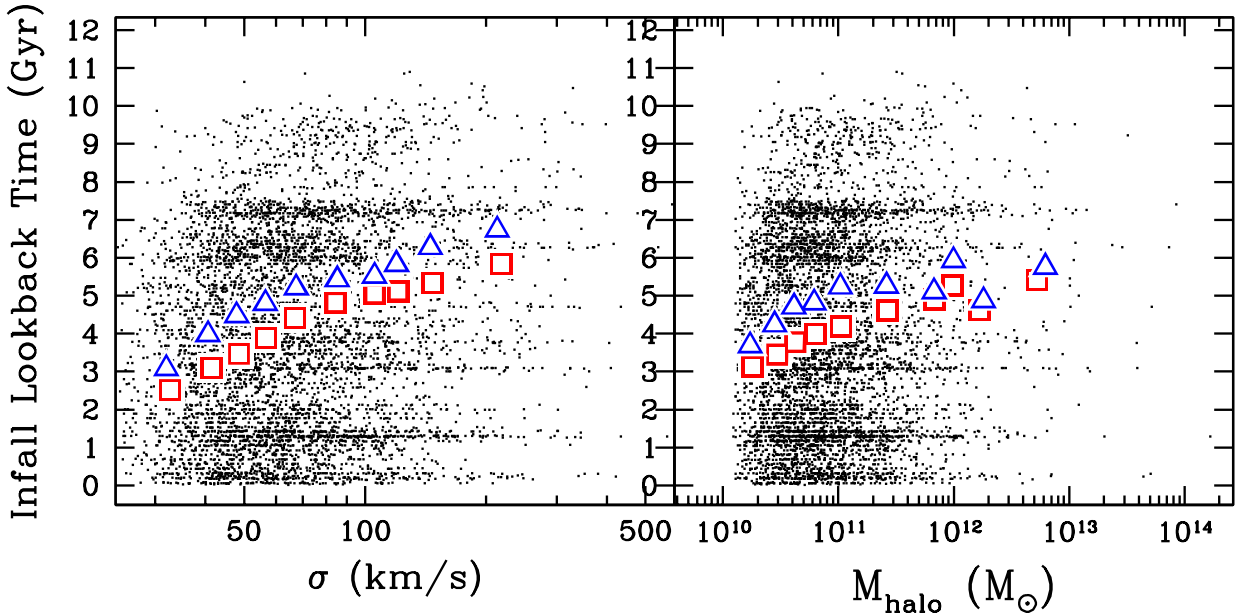}
\caption{The median lookback times at infall of subhalos as a function of pre-infall velocity dispersion (left panel) and pre-infall mass (right panel). The squares show the median infall times of all subhalos which have fallen in. The triangles show median infall times only for those subhaloes which are unlikely to have merged. Halos with higher velocity dispersions are more likely to have fallen in longer ago, whereas this trend is weak for more massive halos.
 \label{fig:tinfall_mvd}}
\end{figure*}

Although there is considerable scatter, subhalo infall times are also correlated with the velocity dispersion and mass of the halo (\figref{tinfall_mvd}). The median infall time for dwarf halos ($\sigma < 50 \kms$) is 2-3 Gyr, while almost all massive subhalos ($\sigma > 150 \kms$) fell in at least 2 Gyr ago with a median of 6 Gyr. This can be interpreted as a consequence of the spatial correlation/clustering of overdensities in the early universe, such that the largest overdensity collapsed into the cluster itself while nearby overdensities formed massive subhalos which fell in at early times. The consequences of this correlation on galaxy formation models are that even a purely environmentally-quenched model would predict a non-zero slope in the colour-magnitude relation, since more massive (and hence brighter) galaxies would have fallen in earlier and hence be redder on average.

\begin{figure}
\includegraphics[width=\columnwidth]{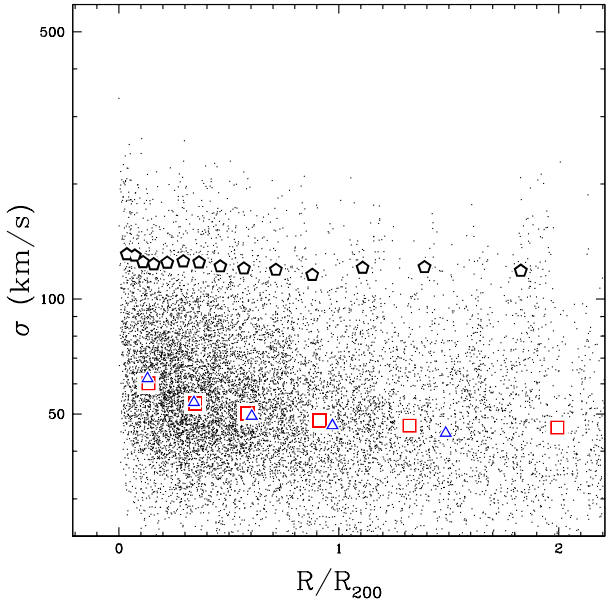}
\caption{Velocity dispersions of halos as a function of projected separation from the cluster centre. Red squares show the median velocity dispersion at various radii; blue triangles exclude orphan halos which are likely to have merged, as in \figref{tinfall_mvd}. Black pentagons show medians only for halos likely to contain bright galaxies. High dispersion subhalos tend to be found near the cluster centre, where the median velocity dispersion is $\sim 20 \kms$ greater than in the field. This trend is not as strong for massive halos likely to host bright galaxies.
\label{fig:position_vd}
}
\end{figure}

In addition to the mass trend, there is a weak but non-zero trend in the median velocity dispersions of halos as a function of cluster-centric distance (\figref{position_vd}), even in projection. This can be viewed as a combination of the infall time versus position (\figref{tinfall_position}) and infall time versus mass relation (\figref{tinfall_mvd}), which essentially causes mass segregation of subhalos. One would then expect a model without any explicit environmental dependence to maintain some age trend, as long as the quenching is tied to halo mass or velocity dispersion. However, the trend appears to be strongest for low-mass halos and is weaker for halos likely to host massive galaxies (which we identify using a typical model from \secref{sfmodels}).

\section{Data}
\label{sec:data}

We constrain our models using data for giant galaxies in clusters from the NOAO Fundamental Plane Survey \citep[hereafter NFPS]{SmiHudNel04} of $\sim 3000$ galaxies in 94 nearby rich clusters.  Colours for 750 giant ($M_R < -20.25$) galaxies in 8 rich clusters are taken from \cite{HudSteSmi10}. This subsample has no colour selection and includes red and blue galaxies.  It provides (total) galaxy colours in $B-R$ as well as colours in the same filters for the bulge and disc components separately.  Spectroscopic data, in the form of central absorption line indices, are from \cite{NelSmiHud05}.  The NFPS spectra are too noisy to obtain ages and metallicities for individual galaxies, but the trends as a function of internal velocity dispersion $\sigma$ and projected distance from the cluster centres are well determined \citep{SmiHudLuc06}.

In addition to NFPS, we also employ a larger data set from SDSS, which remains complete to nearly arbitrary distances from the cluster core. We combine the bulge-disc decompositions of \citet{SimMenPat11} with a catalogue of 625 brightest group and clusters galaxies from \citet{vdLBesKau07}, which is based on the catalogue of \citet{MilNicRei05}. This provides total, bulge and disc colours in $g-r$ for all galaxies in each group, as well as distances relative to the cluster centre (typically the brightest cluster galaxy). Likely background galaxies with projected velocities $v>2\sigma_{3D}$ ($v>3.46\sigma_{1D}$) are excluded, where $\sigma_{1D}$ is the measured (projected) velocity dispersion of the cluster, and $\sigma_{3D}$ the estimated 3D dispersion. To compare to the simulated cluster sample, we select only the 28 clusters with $\sigma_{1D} >$ 800 \kms. Within a distance of 4$R_{200}$, there are about 5,000 galaxies in this sample. This data set will be described further in an upcoming paper (Oman et al., in prep).

\section{Star formation models and colours}
\label{sec:sfmodels}

We will now use the halo orbits of \secref{nbodyresults} as input to simple star formation models in which environmental ``triggers'' quench star formation - more specifically, cluster-centric triggers such as infall past $r_{200}$ or reaching pericentre. This quenching can either be slow (e.g. causing an exponential decay in the star formation rate), or abrupt. 
At first, we will only consider models where quenching occurs within the cluster (i.e. after infall onto the cluster halo); later, we will incorporate pre-processing by allowing quenching by smaller group-sized halos.

\subsection{Galaxy Catalogs and Stellar Masses}

To generate mock galaxy catalogues from the dark-matter only simulation, each subhalo is assumed to host a single galaxy. Stellar masses are assigned to galaxies based on their dark matter halo's velocity dispersion. The favoured model maps dark matter halo (DM) dispersions to stellar (*) dispersions using the relation between maximum halo circular velocity and circular velocity at 10 kpc from \citet{TruKlyPri11}: $\log(\sigma_{*}) = -0.236 \times (\log(\sigma_{DM}))^2 + 1.734 \times \log(\sigma_{DM}) - 0.5723$. Stellar masses are then assigned from a fit to SDSS early-type galaxies \citep{SimMenPat11} using the selection criteria outlined in \citet{TarDubYee13a}: $\log(M_{*})=3.498 \times \log(\sigma_{*})+3.225$. We have also tested an alternative prescription whereby halo velocity dispersions are converted to stellar masses from the best-fit $\sigma-R\sbr{eff}$ relation for Coma galaxies from \citet{AllHudSmi09}, more representative of rich cluster members. While this yields a slightly higher abundance of bright galaxies, the differences in typical galaxy colours between the two prescriptions (0.01-0.02 in B-R) are smaller than typical observational errors on median colours (0.03-0.04). 
Lacking any reason to favour this prediction, we employ the SDSS-based prescription in all cases.

After assigning stellar masses, the star formation history of each galaxy determines stellar mass-to-light ratios and magnitudes. The final simulated galaxy catalogue contains only galaxies brighter than the observational limits of the data, i.e.\ $M_{R} < -20.25$, or a roughly equivalent cut of $M_{r} < -20.01$. These magnitude cuts typically exclude halos with dispersions below $\gtrsim 100$ km/s, or $M\sbr{stellar} \gtrsim 10^{10}  M\sun$ - thus, the catalogue only contains galaxies in halos which are well above the simulation's resolution limit. The subhalos that host these galaxies have masses $M\sbr{halo} \gtrsim 10^{11} M\sun$.  The median and mean stellar masses of galaxies in the sample are $M\sbr{stellar}$ $\sim 2.5\times10^{10}  M\sun$ and $\sim 5\times10^{10}  M\sun$, respectively.

\subsection{Median colours as a function of cluster-centric radius} 
Throughout this section, the primary statistic used to compare models and data will be the median colour in bins of cluster-centric radius. The median has the advantage that it is robust to outliers. 

Although we will discuss qualitatively the colour distributions produced by different models, we are reluctant to draw strong conclusions based on the full shape of the $B-R$ colour distribution for two reasons. First,  the observational uncertainties on individual galaxy colours ($\sim 0.1$ mag for discs; \citet{HudSteSmi10}) are not included in our models; these would scatter the intrinsic distribution and preclude the rejection of any model based on colour bimodality. Second, we will show that our default model for the unquenched blue cloud is simple: a simple exponential timescale prior to infall and a second one following quenching for recently quenched satellites. To realistically model the spread in the blue population, one would likely need to incorporate scatter in the star formation histories and metallicities, none of which are well-constrained by the data.  For these reasons, the median colours are expected to be a more robust statistic than the full distributions of $B-R$ colours. 

The same argument applies to blue fractions. Unlike the case for, say, $u-r$ colours \citep{BalBalNic04}, in $B-R$, \cite{HudSteSmi10} show that there is no well-defined bimodality in the observed colours of cluster galaxies, not in the least because there are fewer blue galaxies than in the field. It is therefore difficult to determine where to separate blue and red. Moreover, given that the blue and red populations are close and overlapping, correct treatment of the scatter  (observational and in the models) is essential to obtaining the correct fractions.

\subsection{Single-component toy models}
\label{sec:singlemodels}

We first explore two simple toy models designed to test the influence of cluster-centric environment. These are ``single component'' models in which the entire stellar population of the galaxy is described by a single star formation history (SFH). Although such models are not expected to be good matches for cluster galaxies on their own, we can test if they are appropriate fits to the distinct bulge and disc components.

\subsubsection{The AHS Star Formation Models}
\label{sec:ahsmodels}

We will use the \citet[hereafter AHS]{AllHudSmi09} SFH models as baselines for our single-component models. The AHS models parametrised the SFH in terms of simple functional forms (e.g.\ single burst, exponential, constant SF followed by ``abrupt quenching'', etc.) and then adjusted the parameters of the model in order to reproduce both the median and the scatter in the global colours and central spectroscopic line indices of NFPS red-sequence galaxies.  Several studies \citep{SmiLucHud09a, GraFabSch09} have shown that stellar populations are more tightly linked to the velocity dispersion than to the stellar mass, so \emph{a priori}, one expects a correlation between mean stellar age and the stellar velocity dispersion.  In the AHS models, the central stellar velocity dispersion determines (1) an age-related parameter, such as a Single/Simple Stellar Population (hereafter SSP) age or a quenching time, (2) the metallicity and (3) the $\alpha$-enhancement. Random scatter is added to each of these parameters to generate a galaxy population with a distribution of ages, metallicities and $\alpha$-enhancements. For each simulated galaxy,  the stellar population parameters are used to generate simulated colours and absorption linestrength indices calculated by convolving simple stellar population (SSP) model SEDs \citep{Mar98, Mar05} and Lick-index absorption linestrengths \citep{ThoMarBen03} with a given SFH in order to reproduce the line indices of red-sequence cluster galaxies from  \cite{NelSmiHud05} and \cite{SmiLucHud07}.  Having done so, AHS find reasonable agreement with galaxy colours as well, with the caveat that $B-R$ colours are too red by about 0.1 mags, a common problem in stellar population models and one which will be addressed shortly. Furthermore, the AHS models include synthetic total magnitudes, allowing us to select samples in the same way as the real NFPS data. Thus by construction the AHS models should match the observed red-sequence stellar population observables, with the caveat that the SSP colours are likely to be too red \citep{MarStrTho09}. Finally, in all cases we assume a universal \cite{Kro01} stellar initial mass function (IMF). 

\subsubsection{Internal Quenching versus Quenching on Cluster Infall}

To test the impact of environment on star formation in cluster galaxies, we employ two simple models - one with an indirect relation between environment and star formation, and one with a more direct impact. The first (``null hypothesis'') model is one in which the cluster environment has no direct effect: we assume that the star formation history depends \emph{only} on internal properties of the subhalo (specifically the velocity dispersion of the subhalo), and not on infall into the cluster. Star formation rates are modelled either as declining until a specified time $t_{AQ}$ or as a single instantaneous burst (SSP). The $\alpha$-enhancement and metallicities are assigned through a lookup table of similar galaxies in the AHS mock catalogues. This yields a red-sequence in which stellar population age, and hence galaxy colour, depends only the \emph{internal} properties and not explicitly on position in the cluster. Nonetheless, colours in this model have a weak and indirect dependence on cluster-centric radius due to the correlation between subhalo velocity dispersion and cluster-centric radius noted in \secref{nbody}. We will explore this trend further in \secref{twocomp}.

In the second simple model, the galaxy's internal velocity dispersion has no effect on its age and metallicity. Instead, the quenching time is set to the lookback time at which it fell into the halo, i.e. the time at which it first crossed $r_{200}$. Metallicities and $\alpha$-enhancements are set to solar with small ($< 0.05$ dex) scatter. We model the star formation rate as exponentially declining - prior to infall, with some pre-infall $e$-folding time $\tau\sbr{pre}$, and subsequently with a post-infall $e$-folding time $\tau\sbr{post}$. We have also considered models where quenching begins at pericentre, where environmental effects are strongest. In practice, most subhalos of the cluster take 0.5 -- 1 Gyr to reach pericentre from $r_{200}$, and so quenching at pericentre is quite similar to (but more physically motivated than) a fixed delay time of about 1 Gyr.

Environmental quenching as in the infall-based model produces a wider range of galaxy colours and a steeper dependence of colour on cluster-centric radius. Galaxies that have recently (or never) fallen in have the bluest colours, dominating the outskirts of the cluster. In the core of the cluster, the time since infall spans a range from several hundred Myr for the most recently accreted halos still on their first orbit around the cluster core, to many Gyr for older, virialised halos. Hence the overall colour gradient is a function of the relationship between infall time and position (\figref{tinfall_position}) modulated by the dependence of colour on infall time. The former relation is almost entirely determined by cosmology, while the latter is dependent entirely on our choice of star formation model.

The simplest models one can construct are either entirely internally quenched, or entirely environmentally quenched. A naive choice for the parameter $\tau\sbr{pre}$ of infinity results in a constant star formation rate; however, this produces excessively blue colours for field galaxies. A similar naive choice of $\tau\sbr{post}$=0 produces excessively large gradients in colours. Finally, the internally-quenched SSP model produces virtually no gradient. These trends are examined in greater detail in Appendix \ref{sec:simplemergers}. We now turn to more realistic bulge/disc models using multiple components based on these simple toy models. 

\subsection{Bulge plus disc models}
\label{sec:twocomp}

\subsubsection{Environmental Dependence of Bulges and discs}

The $B-R$ colours of bulge and disc components of giant galaxies in clusters were studied by \cite{HudSteSmi10}.  They found that bulge colours were redder than discs (as expected) but that bulge colours were independent of cluster-centric radius. In contrast, the colours of disc galaxies depend significantly on cluster-centric radius and drive the weaker gradients observed for global colours. The lack of radial dependence of the bulge colours is therefore similar to the predictions of the ``internally-quenched'' toy model above, whereas the radial dependence of the disc colours is similar to the infall-quenched toy model.  This suggests that we build a composite bulge/disc model in which only the discs are affected by environment.

In the NFPS data, the bulge-to-disc ratio varies along the red sequence, with higher-mass systems being more bulge-dominated. Indeed AHS showed that the bulge-to-total light ratio ($B/T$) was tightly correlated with central velocity dispersion (see their Figure A1). We use their empirical relation to determine $B/T$ at a given velocity dispersion, $\sigma$: $(B/T)_{R} = 0.5\times \log(\sigma_{*})-0.44$. Typical galaxies with velocity dispersions of 100 \kms \ have 54\% of their light in the bulge, while the most massive galaxies are over 80\% bulge-dominated. 

\subsubsection{Bulge Model}
Having assigned bulge fractions, bulge ages are then generated as for the ``internally-quenched'' toy model described above. The bulge metallicity is determined by drawing a random galaxy from the AHS catalogue with a similar stellar dispersion ($\pm 0.1$ dex). The AHS catalogue provides a total metallicity, $\alpha$-enhancement and model-dependent age for the galaxy, derived from fits to line index strengths (not colours) - we use this same age for the bulge, and give the bulge sufficiently large metallicity and $\alpha$-enhancement to match the AHS value. In practice, almost all bulges have super-solar metallicity and are increasingly metal-rich at larger velocity dispersions. By contrast, discs are assumed to have solar metallicity with small ($< 0.05$ dex) scatter and no $\alpha$-enhancement. Since the total metallicity is a mass-weighted average of the disc and bulge metallicities, we adjust the bulge metallicity to reproduce the desired total metallicity (which typically increases the bulge metallicity by 0.05 -- 0.1 dex. The bulge $\alpha$-enhancement is adjusted in a similar fashion.

\begin{figure*}
\includegraphics[width=0.5\textwidth]{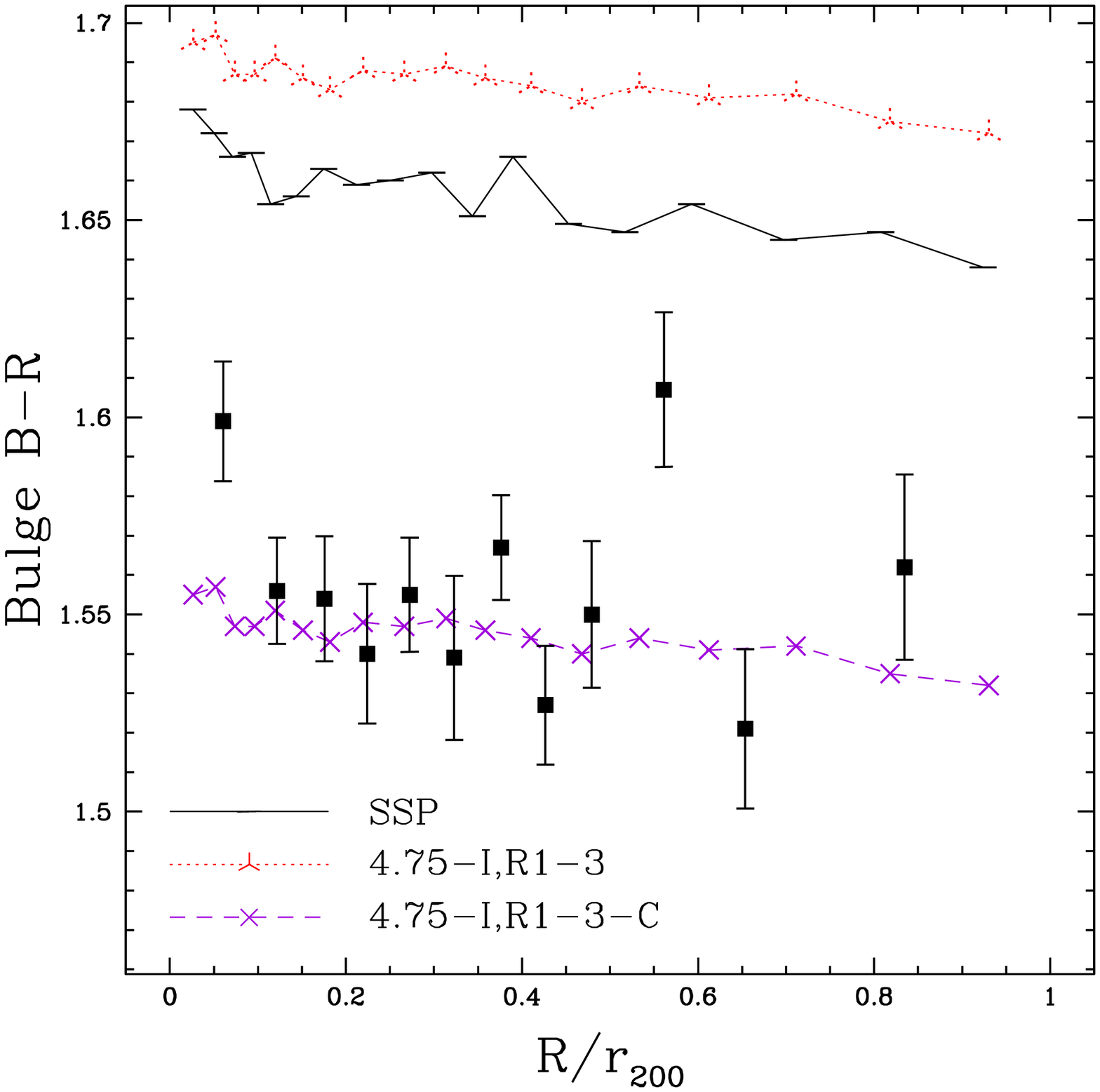}%
\includegraphics[width=0.5\textwidth]{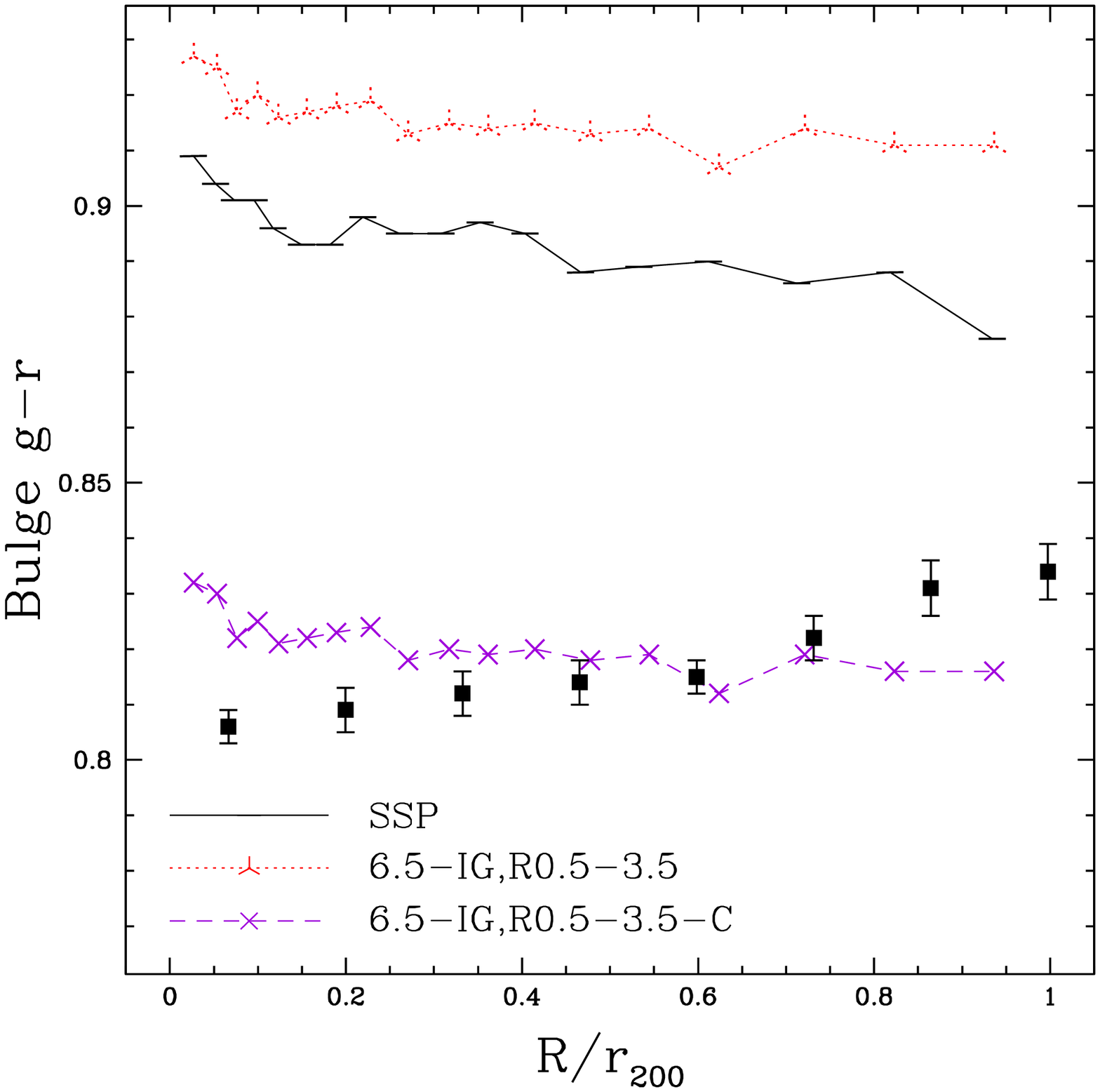}%
\caption{Simulated median bulge $B-R$ and $g-r$ colours for cluster galaxies, compared to data from NFPS $B-R$ colours (left) and SDSS $g-r$ colours (right). The colour of the bulge component is shown for an SSP and a typical bulge + disc model; however, it is flat and essentially independent of the disc model, and remains so well beyond $R_{200}$. However, simulated bulge colours are too red, a common problem in stellar population models of massive early-type galaxies. The corrected model ('-C') is shifted by -0.14 mags (bluer) to better match the observed trend, as discussed in the text. Galaxies with small bulge fractions ($B/T_{R}<$0.2) are excluded from this figure.
\label{fig:colours_bulge}}
\end{figure*}

For the most part, bulge colours are constant with cluster-centric radius in both the models and data, as shown in \figref{colours_bulge}. The main exception is that the SDSS galaxies show slightly bluer bulge colours in the inner regions of the cluster. This could be connected with the fact that bulge fractions in SDSS $r$-band are both lower (about 47\% on average versus 57\% in NFPS) and slightly higher in the cluster core (53\%) than in the outskirts (44\%), a trend which is not present in NFPS. This trend could be real or entirely systematic - the SDSS data covers a much larger redshift range, and was generally taken with shorter exposures and worse seeing on a smaller telescope than NFPS, so the existence of this relatively small trend is not especially worrisome.

Of greater concern is that the simulated bulge colours are too red by at about 0.15 mags in $B-R$ and 0.1 mags in $g-r$. This is a well-known problem in stellar population modelling and is detailed by, amongst others, \citet{MarStrTho09} (whose stellar population models are used here, uncorrected). The AHS models themselves have a similar offset in bulge colours, which can only be partially corrected by solutions such as modelling the effect of $\alpha$-enhancement \citep{AllHudSmi09}. However, it should be noted that the ages and metallicities in the AHS models do reproduce linestrength indices by construction and as intended, likely because the effects of $\alpha$-enhancement are better modelled for linestrength indices. Thus, it is likely that any mismatch in bulge and/or total colours is due to a weakness in stellar population model predictions of colours (but not linestrength indices).

Unfortunately, no current stellar population models entirely correct for this issue. \citet{MarStrTho09} suggest including a small fraction of very metal-poor and old stars; however, little direct evidence exists for such a population, and a very large fraction (about 10\%) is required to entirely explain the offset. $\alpha$-enhancements are not yet accurately or self-consistently modelled. The remaining option is to apply an artificial shift to the colours directly, or to artificially lower bulge metallicity and age. The '-C' (corrected) model in \figref{colours_bulge} applies a fixed -0.14 shift in $B-R$. Alternatively, lowering the metallicity by 0.3 dex and age by 1.5 Gyr accomplishes the same task - however, this is not a suitable physical solution, since these ages and metallicites are inconsistent with ages and metallicities derived from absorption line strengths. Thus, for the remainder of the paper, we will apply this fixed shift of -0.14 in $B-R$ and -0.095 in $g-r$, and models using this shift will be postfixed with '-C'.

\subsubsection{Disc Quenching Model}

The above prescriptions fix the stellar populations of the bulge and also the metallicity and $\alpha$-enhancement of the disc. This leaves only the SFH of the disc to be modelled. We first consider disc colours in the field, where there is assumed to be no quenching.  For field discs, we adopt an exponentially-decaying SFH and constrain the $e$-folding timescale so that the colours match the median colour of discs in the field: $B-R = 1.25 \pm 0.05$ \citep{HudSteSmi10}.  This is in reasonable agreement with the colours found by \cite{MacCouBel04} for discs in field spiral galaxies in the same magnitude range as the NFPS sample. 

We find that the observed colours can be fit with a (pre-quenching) exponential timescale $\tau\sbr{pre} =$ 4.5 -- 5 Gyr. $\tau\sbr{pre} =$ 4.75 Gyr yields a $B-R$ of about 1.26 in the field. Similarly, the SDSS field $g-r$ colour of about 0.6 can be reproduced with $\tau\sbr{pre} =$ 6.5 Gyr. The difference between these two values of $\tau\sbr{pre}$ could again be partly real, or partly systematic differences between bulge-disc decompositions in the two sample, and uncertainties in stellar population models.

In our bulge/disc models, disc quenching will be caused by interaction with the cluster environment. There are two parameters that control the quenching. The first parameter controls how \emph{rapid} the quenching is.  In the simplest scenario, star formation is instantly and completely suppressed, as might be expected if the quenching is due to ram-pressure stripping of cold gas. We refer to this as ``abrupt'' quenching (AQ).  Alternatively, star formation may be quenched more gradually, as in the ``strangulation'' scenario. We model this as a second exponential decay in the star formation rate. The exponential decay time \emph{after} quenching, $\tau\sbr{post}$, is uncertain and so below we experiment with different values. Note that in this context AQ corresponds to $\tau\sbr{post} = 0$. In practice, abrupt quenching is strongly disfavoured by the data, and so non-zero values of $\tau\sbr{post}$ are required.

Two additional parameters control how quenching proceeds. One parameter determines \emph{where} in the cluster quenching first occurs: at $r_{200}$ (I) or at pericentre (P). An additional parameter ``R'' determines how far the influence of the cluster stretches as a factor of $r_{200}$. By default (if not otherwise specified) and in most models this parameter has a value of one - i.e., galaxies are quenched after crossing $r_{200}$. In some models, however, R can be smaller than one. In the case of $R=0.5$, only halos which cross $0.5 r_{200}$ are considered as accreted and hence quenched, so halos with pericentres at $0.6 r_{200}$ would not be quenched.

These three parameters suffice to identify most models. Rather than giving models unique names, we simply refer to them with hyphenated values of these parameters in the format: $\tau\sbr{pre}$-Quenching Location,Radius-$\tau\sbr{post}$. For example, abrupt quenching on crossing $r_{200}$ would be $\infty$-I,R1-0. A more complicated model with $\tau\sbr{pre}$=4 and $\tau\sbr{post}$=3, with galaxies quenched after crossing 0.3$r_{200}$ would be labelled 4-I,R0.3-3.

\begin{figure*}
\includegraphics[width=0.5\textwidth]{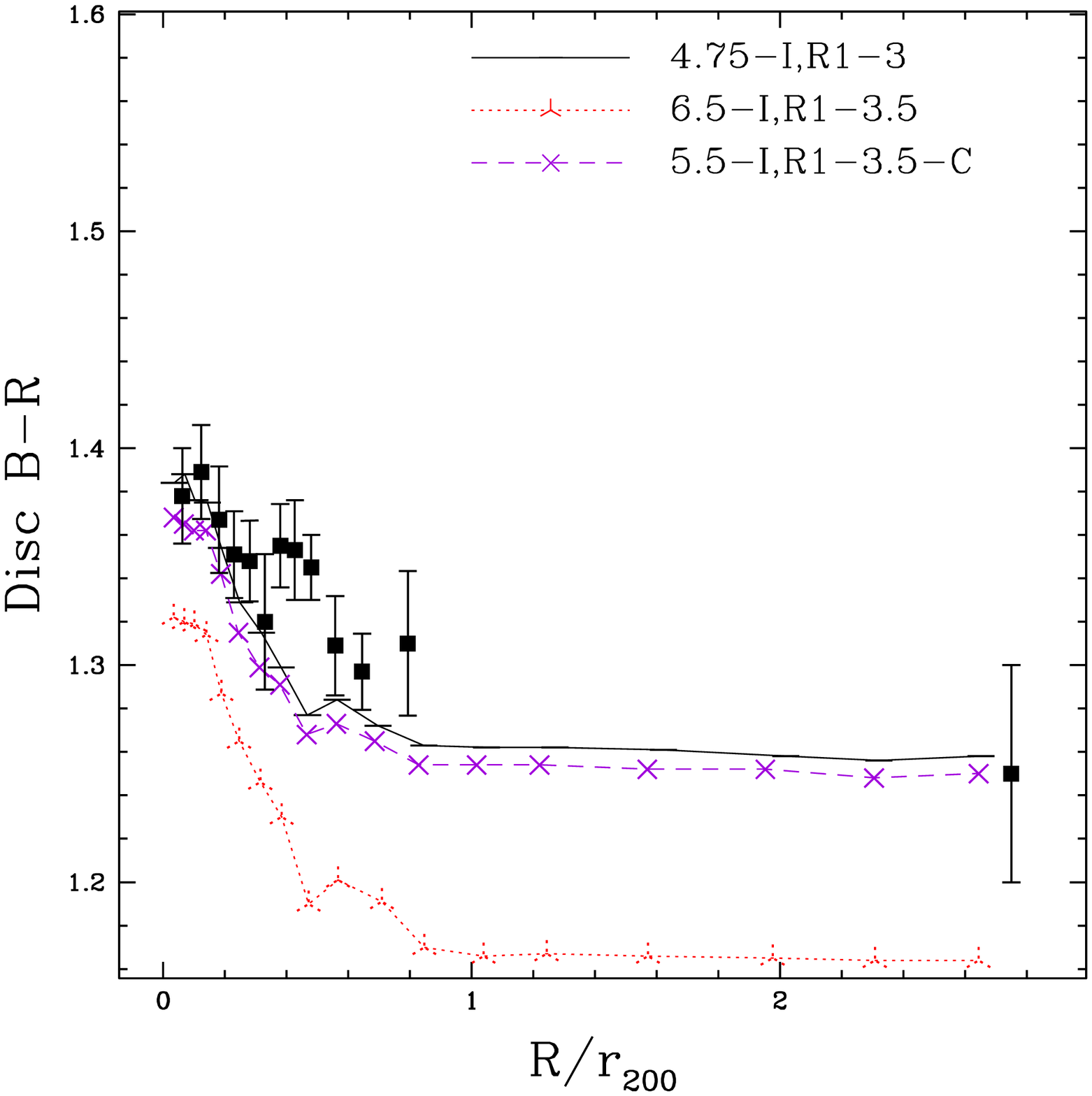}%
\includegraphics[width=0.5\textwidth]{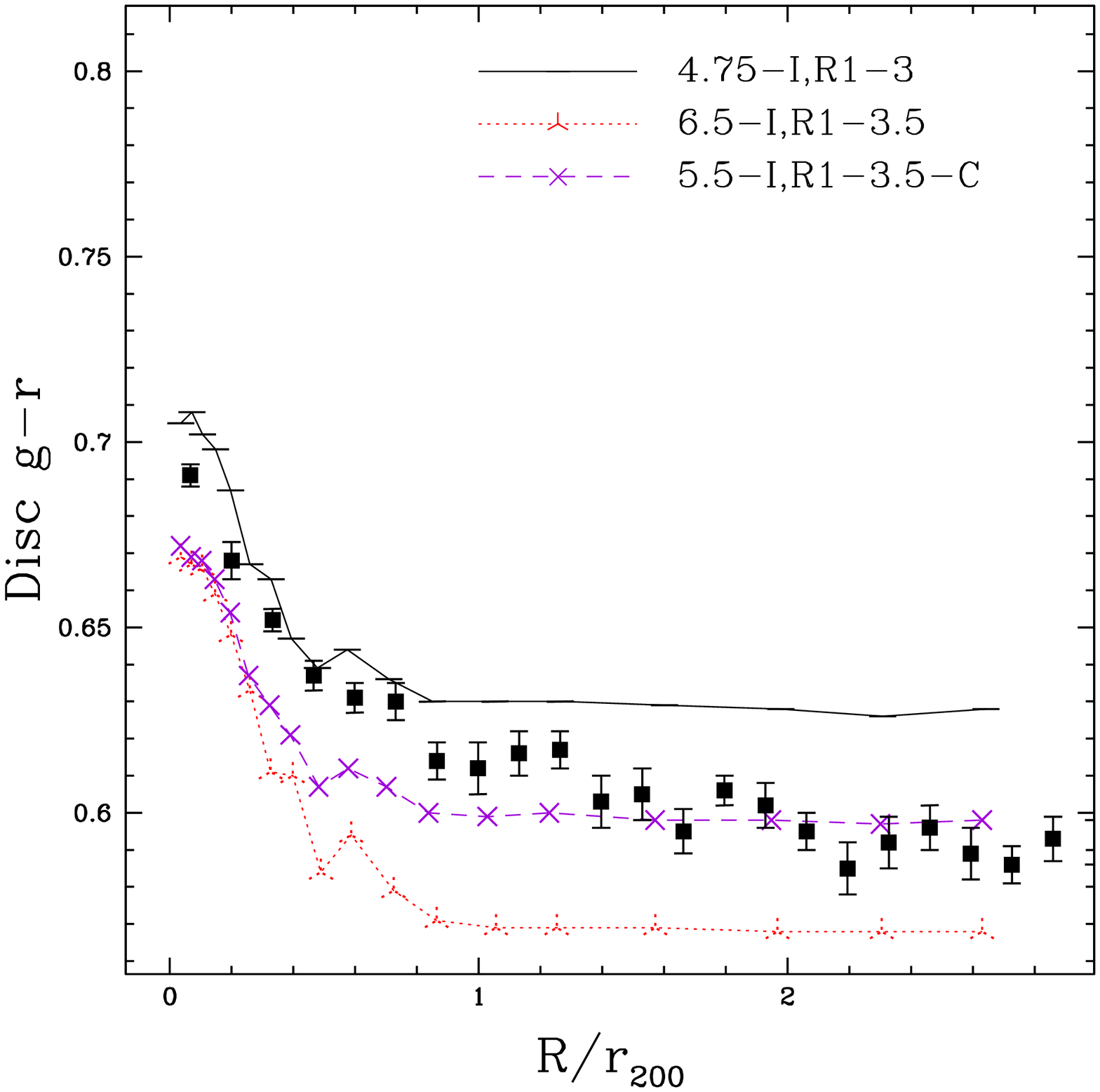}%
\caption{Simulated median disc $B-R$ and $g-r$ colours for cluster galaxies, compared to data from NFPS and SDSS (as in \figref{colours_bulge}). These models incorporate merging but not pre-processing, with a $\tau\sbr{pre}$ tuned to fit field colours in each sample. The SDSS data prefer larger values of $\tau\sbr{pre}$ and $\tau\sbr{post}$. In both cases, the trend in median colours within $0.5R/r_{200}$ is too steep. The outermost data point for NFPS represents the colours of field galaxies and is not actually a bin at 2.7$R/r_{200}$. The '-C' (corrected) model represents an ``average'' model, which fits both NFPS and SDSS data with small offsets in the $B$ magnitude of the disc: +0.04 in $B$, and -0.02 in $g$. Galaxies with large bulge fractions ($B/T_{R}>$0.8) are excluded from this figure.
\label{fig:colours_disk_nopre}}
\end{figure*}

\figref{colours_disk_nopre} shows the best-fit models incorporating merging but not pre-processing, for both NFPS and SDSS data. Again, a $\tau\sbr{pre}$ value of 4.75 Gyr fits NFPS data well, while the colours of SDSS galaxies at $R/r_{200}>$2 are best fit by a value of 6 -- 6.25 Gyr. Similarly, the best-fit values of $\tau\sbr{post}$ are 3 Gyr for NFPS and 3.5 Gyr for SDSS. As previously mentioned, it is not known to what extent these offsets are due to systematics in the observations or models. However, we do note that an intermediate model with $\tau\sbr{pre}$=5.5 Gyr and $\tau\sbr{post}$=3.5 Gyr needs only small corrections of order 0.03 mags in either direction (+0.04 in $B$, -0.02 in $g$) to reproduce both colours, and such errors are rather small considering known systematics in stellar population modelling and comparing two entirely different samples taken with different filters on different telescopes. For example, an offset of 0.15 in $g$ and 0.02 in $g-r$ exists between a Salpeter IMF, $\tau$=1Gyr model with \cite{Mar05} population models compared to the updated \cite{BruCha03}.

What is clear that some quenching is required to reproduce the colour trends, in the sense that $\tau\sbr{post}$ is significantly smaller than $\tau\sbr{pre}$ - models with $\tau\sbr{post}$=$\tau\sbr{pre}$ do not produce any colour trend. However, the quenching appears mild on average - $\tau\sbr{post}$ is smaller than $\tau\sbr{pre}$ by about 40\% across most models, but $\tau\sbr{post}$ - which is essentially the quenching timescale - is still quite long, at between 3 -- 3.5 Gyr. It is also evident that models without pre-processing produce too steep of a trend in colours, even if the quenching radius is extended to 1.5$r_{200}$. Although the median colour at the cluster core is matched, the model colours drop to the field value at about 0.5$R_{200}$, whereas observed colours only appear to flatten well beyond $R_{200}$. This is simply because no galaxies are quenched prior to crossing $r_{200}$, and the median colours at 0.5 -- 1 $R_{200}$ are still sensitive to projected field galaxies, none of which are quenched in this model. Thus, pre-processing appears necessary to match the observed shallow slope in disc colours, regardless of the value of $\tau\sbr{pre}$.

\subsubsection{Pre-Processing: Quenching in groups before cluster infall}
\label{sec:preprocessing}

Although rich clusters are extreme environments and a likely location for environmental quenching, it is also possible for mechanisms such as stripping and strangulation to occur in less massive clusters and groups prior to infall onto the richest clusters - so-called ``pre-processing''. To implement pre-processing, we allow any halos of mass greater than $10^{13}M\sun$ to quench infalling satellites. Models incorporating such pre-processing are labeled with a ``G'' following $\tau\sbr{post}$, since infall can now occur onto group-sized halos.

Pre-processing quenches cluster members prior to their infall onto the cluster, increasing their (lookback) infall time, and also quenches field galaxies which have yet to fall in to the cluster. For typical models, 60-65\% of cluster members are quenched in a smaller group prior to infall; a slightly higher fraction of projection effects are also quenched. These fractions are larger than the 40-50\% found by \citet{McGBalBow09}, but note that the clusters in our sample are also more massive than the most massive of \citet{McGBalBow09}. \citet{McGBalBow09} also found that pre-processed fractions are larger for more massive clusters.

Pre-processing can occur any time from 0-10 Gyr before infall onto the cluster, but median and mean times between pre-processing and infall onto the cluster are 2.5 and 3.5 Gyr, respectively. Thus, pre-processed cluster members are quenched about 3 Gyr earlier than they would have been without pre-processing. Finally, close to half of all galaxies in the simulation have been quenched, so about half of the field has been pre-processed. This change in the fraction of quenched galaxies at all radii - even well outside $R_{200}$ - can be compensated for by increasing $\tau\sbr{pre}$. In practice, SDSS field colours remain virtually unchanged with pre-processing, and only a very minor change from 4.75 to 5 Gyr is needed for the NFPS data.

\subsubsection{Delayed Quenching and Bimodality}
\label{sec:delayed}

In a similar study to this one, \cite{WetTinCon13} found good agreement with SDSS observations of specific star formation rates (sSFRs) using models with ``delayed-then-rapid'' quenching.  The rapid quenching preserves the bimodality in the sSFRs observed by \cite{WetTinCon12}, whereas the delay is necessary to match the overall quenched fraction as a function of environment. Their best fit model does not begin quenching until $\sim 3$ Gyr after infall, whereupon quenching is rapid: $\tau\sbr{post} \sim 0.5-1$ Gyr. We implement this by adding a constant delay time to the quenching time. The effect is similar to pericentre-based quenching, but with a longer delay after infall: pericentre occurs after at most $\sim 1$ Gyr. After 3 Gyr, galaxies are typically in the ``backsplash'' population (or soon to be virialized).

\begin{figure*}
\includegraphics[width=0.49\textwidth]{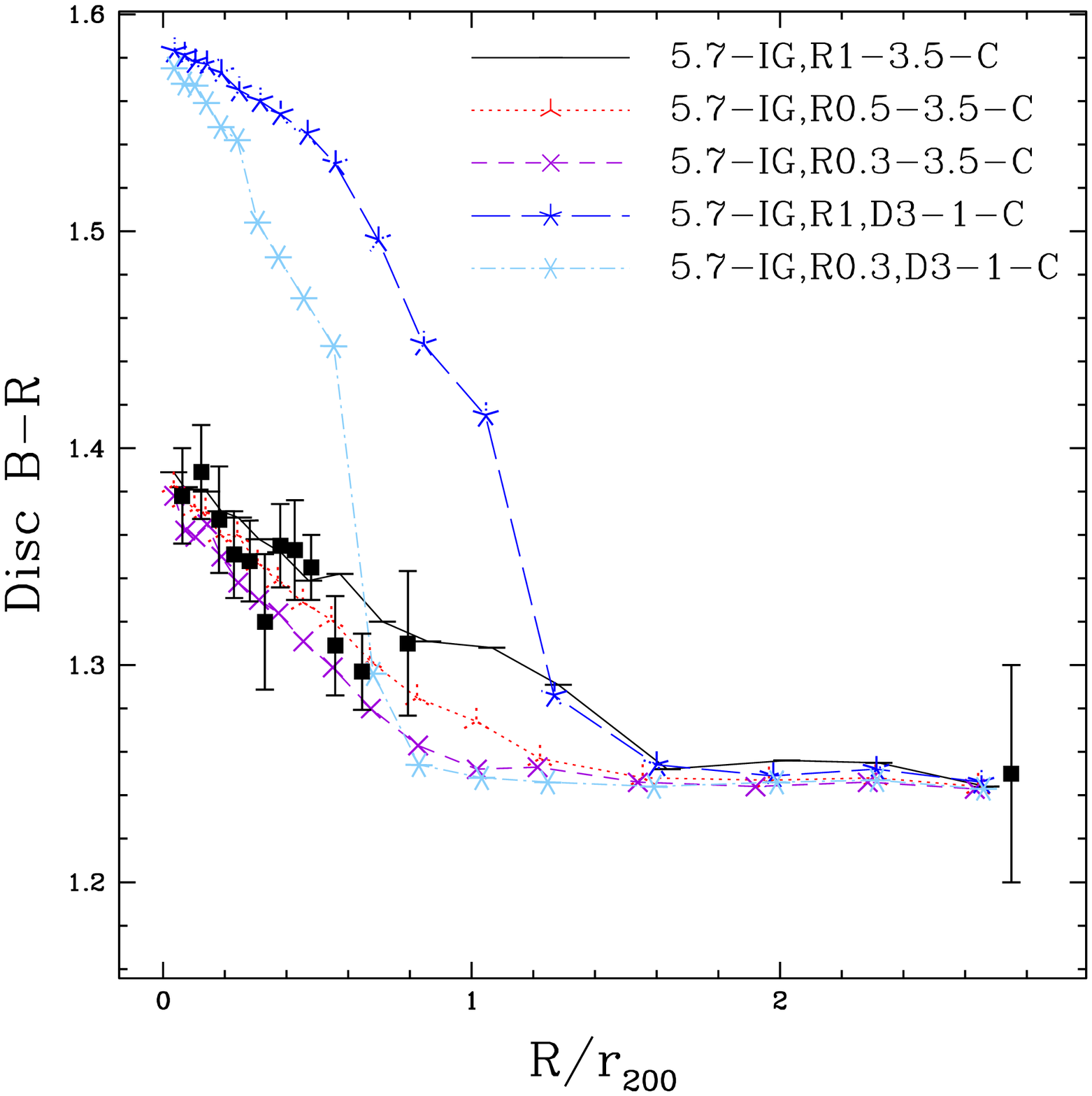}
\includegraphics[width=0.49\textwidth]{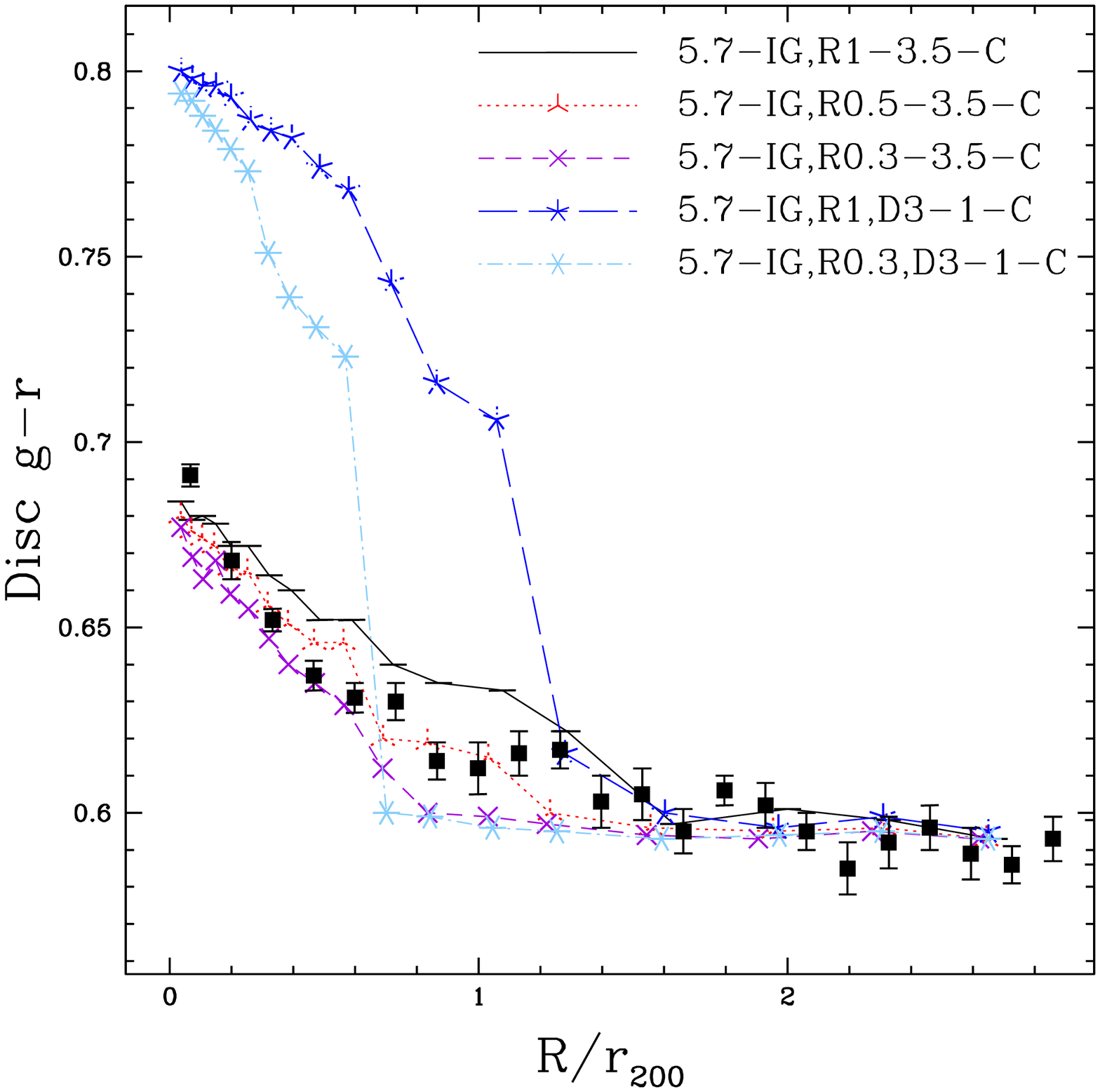}
\caption{As in \figref{colours_disk_nopre}, but now for models with pre-processing in groups as well as the main cluster halo. Models are also shown with quenching at radii smaller than $r_{200}$, or with a delay time D between infall and quenching. The best-fitting model only quenches galaxies falling within 0.5$r_{200}$, whereas the ``delayed-then-rapid'' quenching model produces excessively red discs near the cluster center. All disc colours are corrected ('-C') to match median trends, and galaxies with large bulge fractions ($B/T_{R}>$0.8) are excluded, as in \figref{colours_disk_nopre}. 
\label{fig:colours_disk_pre}}
\end{figure*}

\figref{colours_disk_pre} compares this ``delayed-then-rapid'' quenching model to three best-fit strangulation models. The two 3 Gyr delay model (labelled with ``,D3'', since $\tau\sbr{post}$ now only applies after this delay is completed) uses $\tau\sbr{post} = 1$, which is towards the high end of the range for massive galaxies considered by \cite{WetTinCon13}. As expected the additional delay keeps more galaxies on the ``blue cloud'' compared to a strangulation model with the same $\tau\sbr{post}$. But a weakness of these models is that the mean slope of the $B-R$ vs $R/r_{200}$ is $\sim 0.25-0.3$ magnitudes, whereas the data favour a shallower slope $0.1\pm0.025$ \citep{HudSteSmi10}.  This steep slope is found for \emph{any} model with $\tau\sbr{post} < 1$, whether delayed or not, and the mismatch with observations grows worse with the even smaller values of $\tau\sbr{post}$ favour by \cite{WetTinCon13} for massive galaxies. Moreover, the galaxies closest to the cluster have disc $B-R$ colours that are \emph{redder} than the corrected (and observed) bulge colours. While the disc is younger than the bulge and bluer in pre-correction $B-R$, it is only bluer by a smaller amount than the offset between predicted and observed bulge colours - an uncomfortably thin margin. 

One alternative to delayed quenching is simply quenching fewer galaxies by tightening the criteria for quenching. We also tested models quenching only galaxies falling within 0.5 or 0.3 $r_{200}$. In fact, a quenching radius of 0.5 $r_{200}$ is a better fit to SDSS data and no worse a fit to NFPS data than quenching within $r_{200}$. This model also lessens the tension in colour bimodality, because at any given time there are fewer galaxies undergoing quenching, and hence fewer intermediate galaxies in the ``green valley''. Applying a similar change to the quenching radius does not, however, improve the delayed-then-rapid quenching models, which maintain the same excessively large slope in disc colours over an even smaller spatial range.

\subsection{Consistency check: total colours}

\begin{figure*}
\includegraphics[width=0.49\textwidth]{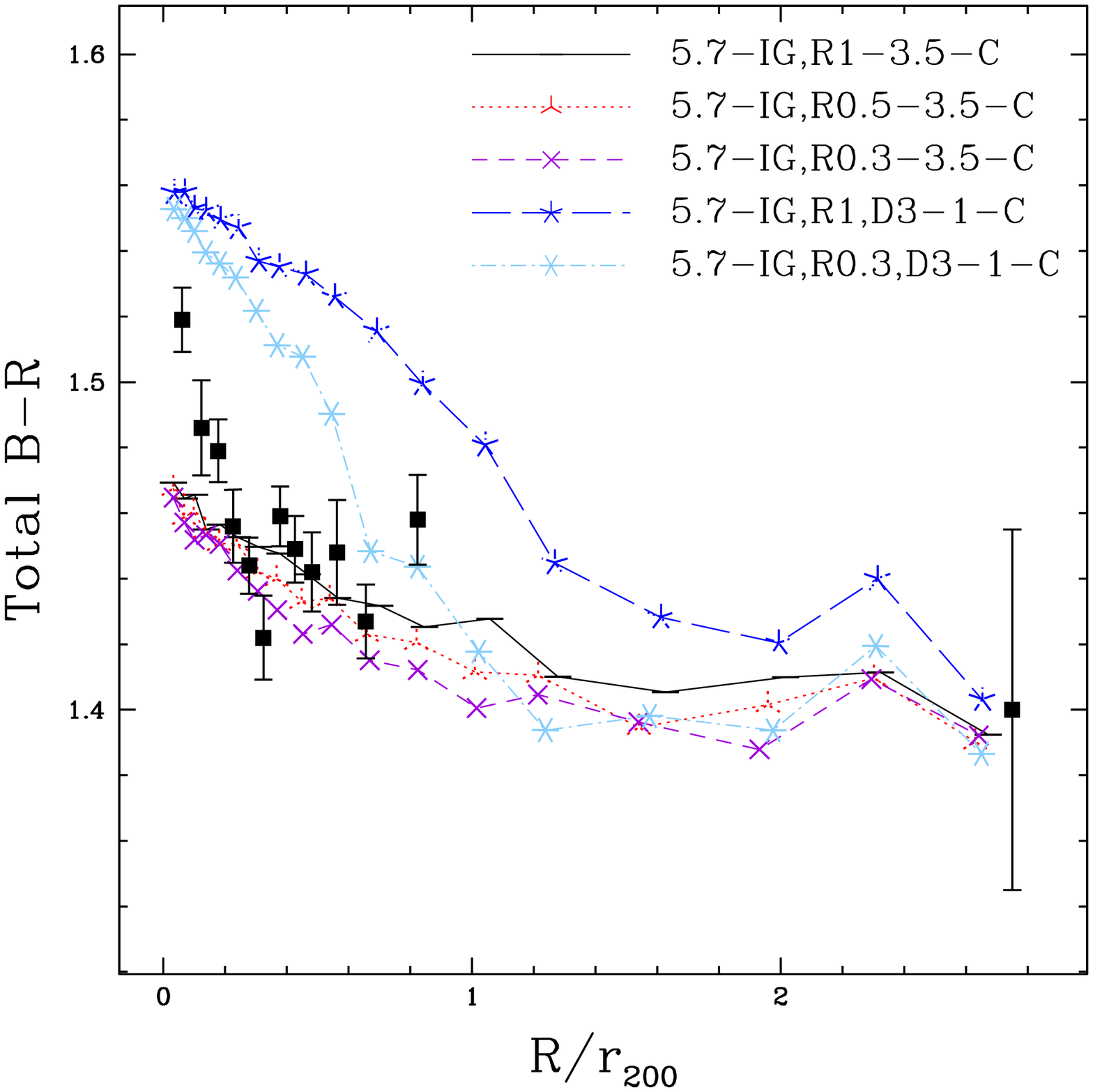}%
\includegraphics[width=0.49\textwidth]{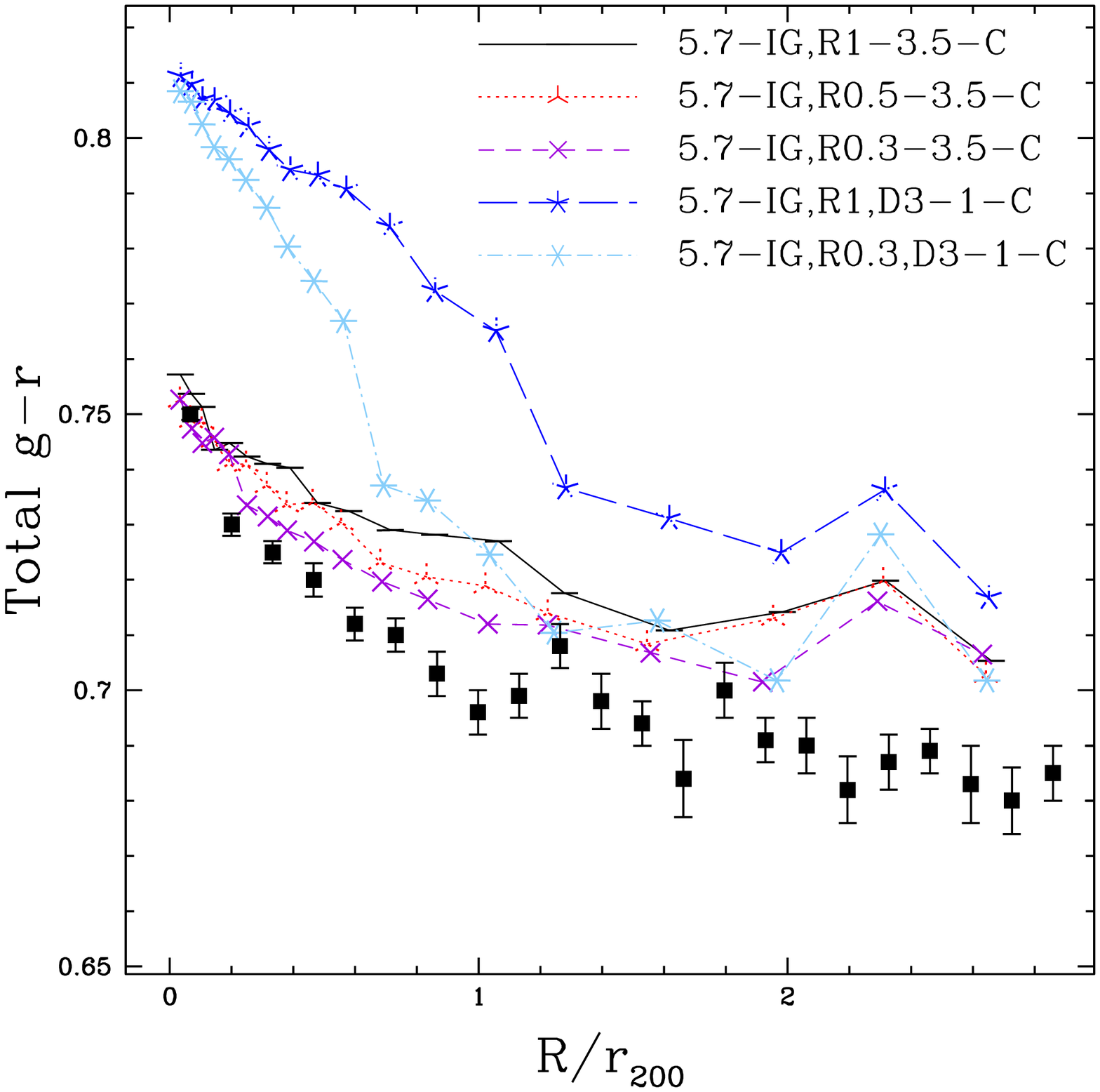}%
\caption{As in \figref{colours_disk_pre}, but now for total colours rather than disc colours. All models except the delayed-then-rapid quenching models produce similar (but slightly shallower) slopes and appropriate intercepts, after correcting for excessive reddening in the bulge. Observed colours are redder than the models in the innermost regions, as observed galaxies have higher $B/T$ close to the cluster core, whereas this trend is much weaker in the models. The fact that the trends are similar in slope (but weaker overall) than those for disc colours in \figref{colours_disk_pre} demonstrates that the total colours are consistent, but also that disc colours provide superior constraints on quenching models, being independent of bulge fractions.
\label{fig:models_pre_br}}
\end{figure*}

We now construct bulge plus disc models and compare them to total colours. In all cases, the bulge component is an SSP; only the disc SFH varies. Such models all produce colour gradients which are mostly compatible with the data (\figref{models_pre_br}). The slopes are somewhat shallower than observed in all models except the delayed-then-rapid model. Differences between colours in the field and at the center of the cluster are about 0.12 in $B-R$ (albeit very uncertain due to large errors on field galaxy colours) and 0.065 in $g-r$, but closer to 0.15 in $B-R$ and 0.09 in $g-r$ for the delayed-then-rapid model, and about 0.07 in $B-R$ and 0.04 in $g-r$ for the remaining models. These shallower slopes are a better fit to the shape of both relations, excepting only the innermost, very red bin in both NFPS and SDSS.  At very small radii (within $0.2r_{200}$), bulge fractions in SDSS and NFPS are higher than in other spatial bins by 5 -- 10\%. In NFPS, this is offset by high bulge fractions in the outermost bin, and so the overall trend in $B/T_{R}$ is flat. The total colours are redder than the model in exactly these two inner- and outer-most bins, where the model $B/T_{R}$ is lower than in the data. In SDSS, there is a radially-dependent trend in $B/T_{r}$, which is not reproduced in the model. $B/T_{r}$ is also slightly and systematically smaller than $B/T_{R}$ by another 5 -- 10\%, and so the model colours are redder than the data further from the cluster centre.

In conclusion, the total colours are entirely consistent with the disc and bulge colours shown before, with shallower model slopes only due to a weak increase in $B/T_{r}$ in the data, which is not modelled. Despite the fact that the uncertainties on total colours are smaller, model total colours are relatively less sensitive to the quenching model because they have large bulge fractions but nearly constant bulge colours. Disc colours are a better discriminant of star formation models. This is true not just for the data and models compared here, but for any sample of galaxies with substantial old and red passive components.

\subsection{The effect of dust}

Galaxy colours are, of course, sensitive to the presence of dust.  \cite{DriPopTuf07} find that, on average, bulge+disc galaxies in the field have a some attenuation of $B$-band light, well-described by a central face-on opacity $\tau\sbr{B}^f \sim 4$ in the models of \cite{TufPopVol04}.  It is well known that spiral galaxies in clusters are stripped of HI \citep{SolManGar01}.  If dust is also stripped, then this would introduce a difference in colour solely due to the absence of dust in the clusters compared to its presence in field galaxies. Using the models of  \cite{TufPopVol04}, we estimate that if all the dust was completely stripped, the effect on disc colour would be to make the discs bluer by 0.10 to 0.22 magnitudes in $B-R$ (with a median of $0.14$ for a typical disc with an inclination $i = 60 \degr$).  Moreover, if the discs on the outskirts were not stripped (and hence contained dust and were redder) and the ones in the centre were stripped (and hence bluer), then the effect would be to make observed gradients shallower. Thus it is possible that the intrinsic (dust-free) cluster-centric gradient in disc colour is considerably steeper than the observed one.

There are, however, two arguments against such an extreme scenario. First, according to the models of  \cite{TufPopVol04}, the bulge light (and bulge colour) are also strongly affected by dust. If the dust is stripped then the bulge will also appear bluer, and more so for bulges than for discs.  In their models, for a typical galaxy with a disc inclination of $i = 60\degr$ the bulge is reddened by 0.35 mag. Thus, if stripping were important, we would expect bulges in the cluster core to be bluer than those at the cluster edges by 0.35 mag. Yet \cite{HudSteSmi10} show that there is no gradient in bulge colour as a function of cluster-centric radius in NFPS data; SDSS bulges are slightly bluer near the cluster core, but the trend is weak. While it remains possible in principle that an intrinsic dependence of bulge stellar populations exactly cancels the putative ``dust removal'' gradient, this seems rather contrived. Rather it suggests that the dust is not stripped in the cluster core. 

Second, it is also possible to study dust stripping directly. \cite{CorDavPoh10} studied dust stripping in Virgo. They found that there was evidence of dust stripping from the outer edges of discs, but the amount of dust stripping was high only in the most extreme ``HI-deficient'' galaxies (defHI $> 1$). Galaxies with such high levels of HI-deficiency are very rare in rich clusters, comprising only a small fraction of the spiral galaxy population \citep{SolManGar01}. We conclude that there is no strong evidence for dust stripping due to the cluster environment in the bulk of the infalling spiral population. Nonetheless, we can also address this concern by comparing model predictions with measures less sensitive to dust properties, such as line index gradients.

\section{Comparison with stellar absorption line indices}
\label{sec:lineindices}

Stellar absorption lines can yield information on the ages and metallicities of non-star-forming stellar populations. In particular the strength of the Balmer absorption lines are sensitive to stellar populations that are $\sim$ 0.5-2 Gyr old, where colours may not discriminate well. In this section, we will compare the models to the NFPS data for red (i.e. non-starforming) galaxies only. Note that the NFPS spectra are fibre-spectra of the central regions (2 arcsecond diameter fibres). We can predict how much of the light within this central aperture arises from the bulge, and how much is from the disc, by using our bulge/disc model. This calculation has been done by \citet[see their Appendix A1]{HudSteSmi10} who derive an ``aperture'' $B/T$ as a function of $\sigma$, which ranges from 75-100 percent (depending on velocity dispersion) for the NFPS sample. We adopt their empirical scaling here: $\sigma$: $B/T=0.5\times \log(\sigma_{*})-0.24$, with the same slope as for $B/T_{R}$ but a larger intercept, so that 100 \kms dispersion galaxies have 76\% bulge light and 300 \kms giants are entirely bulge dominated within the fibre aperture. Given the fraction of light in the fibre from the bulge and from the disc, we simulate the central spectral line indices observed by NFPS.

The NFPS line index data are for red galaxies only. To exclude the blue cloud in our models, we reject model galaxies bluer than 0.2 magnitudes from the red sequence, which is assumed to have a slope of 0.05/mag. In practice, the line index strengths themselves are only weakly dependent on bulge fraction prescriptions (since fiber bulge fractions are so large). However, line strength trends are sensitive to the photometric bulge fraction prescriptions. Photometric bulge fractions are typically only 50 -- 60\%, and so total colours and hence the colour selection are sensitive to both the disc and bulge models and the photometric bulge fractions in both passbands ($B$ and $R$).

\begin{figure*}
\includegraphics[width=0.49\textwidth]{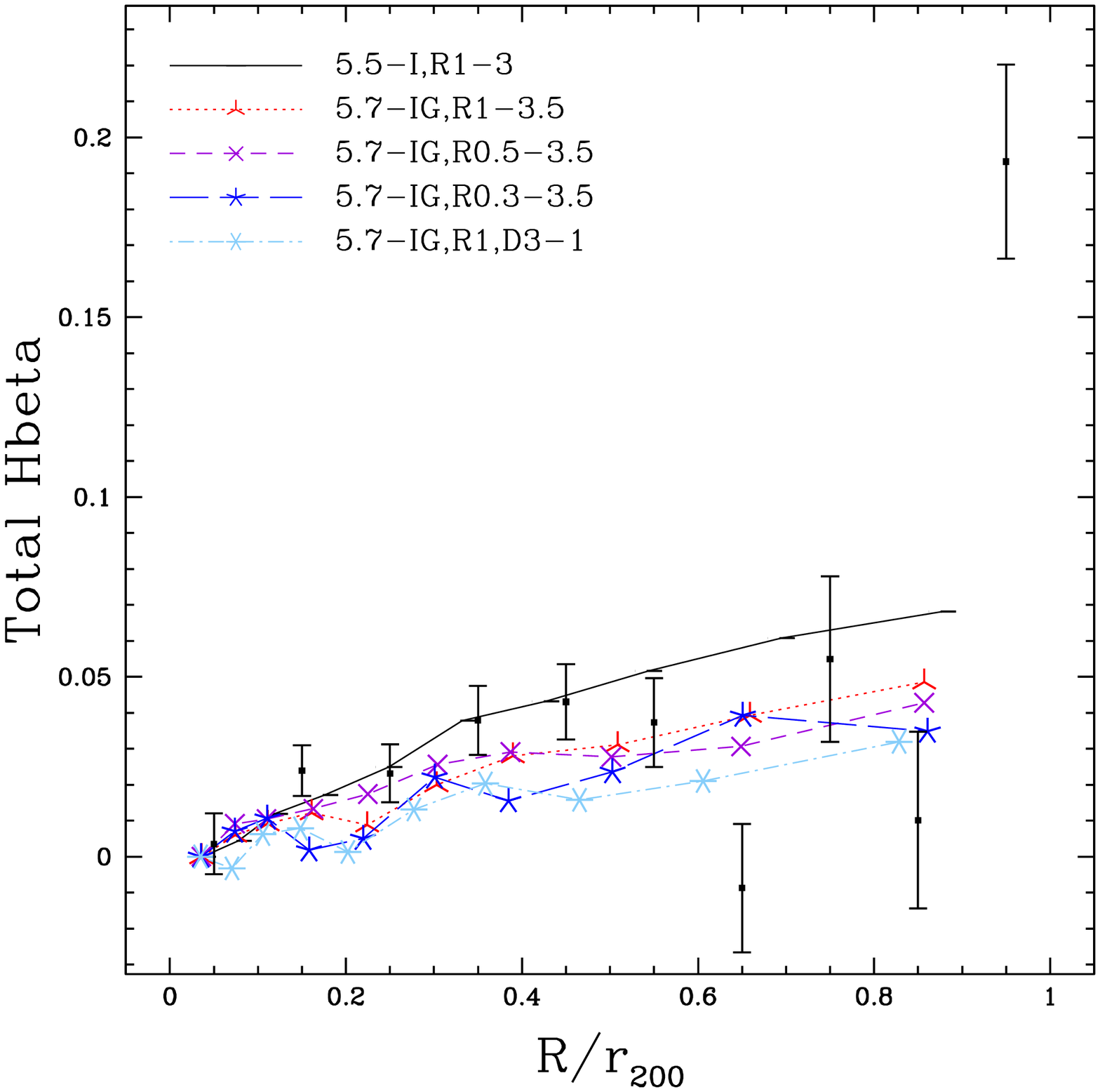}%
\includegraphics[width=0.49\textwidth]{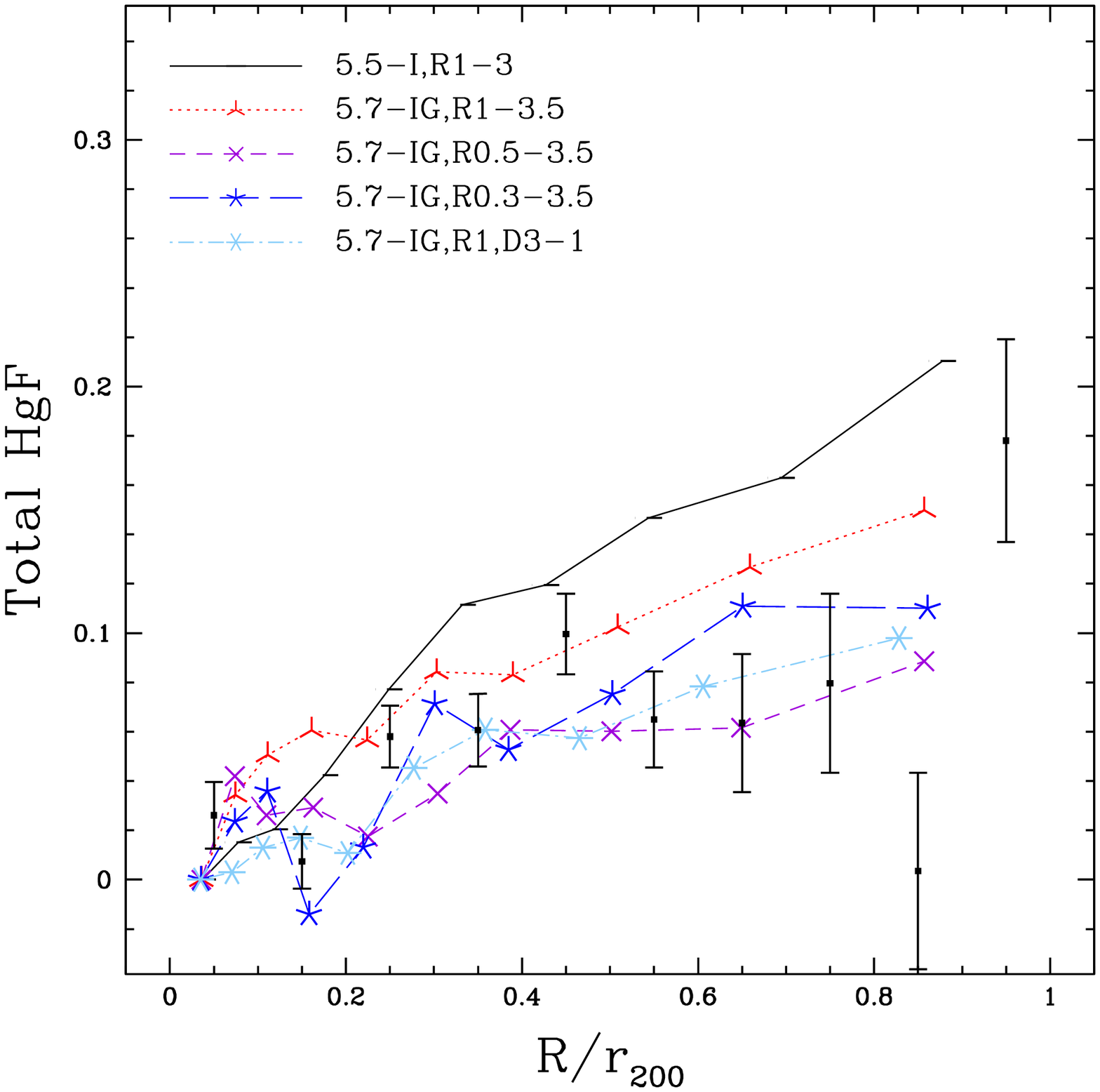}%
\caption{Balmer linestrength index gradients for bulge+disc models, including models with group pre-processing. Virtually all models are consistent with the H$\beta$ and H$\gamma$ gradients, given the large error bars. Most models produce very similar slopes, except for the model without pre-processing, which shows the strongest trends.
\label{fig:models_pre_lines}}
\end{figure*}

The relation between projected position and median linestrength indices is shown in \figref{models_pre_lines}. Most of the best-fitting disc models generate comparable gradients. The colour cut - galaxies bluer than 0.2 from the red sequence - is necessary to exclude star-forming galaxies with emission filling in stellar absorption lines. Unfortunately, this cut also only leaves measurable trends for galaxies nearest to the red sequence, which tends to be similar in most models. The ``delayed-then-rapid'' quenching model produces the weakest H$\beta$ trend, but both trends are not significantly weaker than for the preferred 5.7-IG,R1-3.5 model. Only the model without pre-processing creates a distinct trend, having the largest slope in both lines - still, only the H$\gamma$ trend appears significantly steeper than the data. Thus, like total colours, the line indices do not appear to be a strong constraint on disc quenching models, which is perhaps not surprising given the large fiber bulge fractions and the necessary exclusion of star-forming galaxies.


\section{Discussion}
\label{sec:discuss}

We have shown that models with rapid quenching fail to match the observations. Models with ``abrupt'' quenching,  corresponding to $\tau\sbr{post} = 0$, produce discs which are redder than observed, and a cluster-centric radial gradient in disc colours which is too steep (see Appendix \ref{sec:simplemergers} for full details). Models in which star formation is quenched more slowly (3  -- 3.5 Gyr) produce shallower disc colour cluster-centric radial gradients and slightly steeper absorption line gradients, in accordance with observations (\figref{colours_disk_nopre}). We have also found that models incorporating pre-processing in groups of total mass $M>10^{13}M\sun$ more accurately reproduce the shallow slope of disc colour trends (\figref{colours_disk_pre}).

Several recent studies \citep{WeiKauvon09, WetTinCon13} have attempted to reproduce the observed bimodality in specific star formation rates (sSFRs) in SDSS galaxies with (respectively) full semi-analytic models or simple parametric star formation models. The generally-favoured model with $\tau\sbr{post} \sim 3.5$ Gyr does not produce a strong bimodality in disc or global galaxy colours as a result of the relatively long $e$-folding time. 
For this reason, the models of \cite{WetTinCon13} prefer a short quenching $e$-folding time, equivalent to our $\tau\sbr{post}$. To avoid producing too many red galaxies, they introduce a $\sim 3$ Gyr delay before quenching begins. 

However, \figref{colours_disk_pre} shows that models with a delay time and a short $\tau\sbr{post}$ also produce discs in cluster centres that are too red compared to those at the virial radius. As an alternative to the quenching delay, we presented models where quenching occurs only within a smaller radius, e.g. 0.3 or 0.5$R_{200}$. These models allow a greater fraction of galaxies to remain un-quenched (and hence blue) for longer, because some galaxies will have orbits whose pericentres fall outside the quenching radius on first passage but which, due to dynamical friction, will only pass within the quenching radius on a later pericentre. \figref{tinfall_position} suggests that second pericentric passages occur 2--5 Gyrs after infall, similar to the 3 Gyr delay required by \cite{WetTinCon13}. 

One physically motivated alternative to the delay time is to apply quenching at pericentre, where the intra-cluster medium is densest and tidal forces may be strongest, instead of immediately after infall. Typical infall times from $r_{200}$ to pericentre in the main cluster halo are at most 1 Gyr, and can be shorter at early times. We find that this delay has a very modest impact on galaxy colours. Moreover, if the quenching radius is smaller than $r_{200}$, the delay time will become even shorter. Thus, it is generally difficult to constrain exactly where quenching occurs. If pre-processing is important, then satellites are unlikely to be significantly quenched much further than at $r_{200}$.

We can also compare the quenching times in our best-fit models to those quoted in other recent works employing environmental quenching. The best-fit $\tau\sbr{post}$ is generally 3--3.5 Gyr for models without a delay, with or without pre-processing. \cite{WanLiKau07} modelled quenching of satellite galaxies, favouring an $e$-folding time (i.e. $\tau\sbr{post}$) of 2.5 Gyr, similar to our own best fits. \citet{DeWeiPog11} presented models suggesting that satellites on average spend 5-7 Gyr within a halo of mass $M > 10^{13} M\sun$ before being quenched. This 5-7 Gyr quenching time is not entirely inconsistent with some models depending on when one considers a galaxy to be quenched. For example, in $\tau\sbr{post}=3$ Gyr models, it would take nearly 7 Gyr for star formation to drop to 10 per cent of its pre-quenching value after reaching pericentre.

There are several ways in which these studies could be improved. The overmerging problem can be addressed partially with better numerical resolution. More recent halo finding algorithms use particle velocities in addition to positions and binding energies \citep{OmaHudBeh12}, which can improve estimates of infall and pericentre times and pericentre distances and obviate the need for ad-hoc merging prescriptions. Halo velocities can also be used to disentangle projection effects and backsplash galaxies \citep{MahMamRay11, OmaHudBeh12}, though more observations would be required to test that prediction.

Because of our focus on reproducing mean/median gradients, we did not include many ingredients into our disc models, such as mass-dependent SFRs, and scatter in the SFHs which one might model as variable $\tau\sbr{pre}$. These ingredients would likely help in reproducing the scatter about the mean relations and better reproducing mass-dependent trends, e.g. for dwarfs vs.\ giants, for which we do note substantial differences in the infall history. Another missing ingredient in the models is variable or density-dependent quenching strengths. Quenching mechanisms like ram-pressure stripping can depend strongly on intra-cluster gas density, pressure and satellite infall velocity\citep{GunGot72, McCFreFon08}. The inclination of the satellite's disc may also have an impact. These effects could introduce significant scatter in the values of $\tau\sbr{post}$ and explain the more extreme observations of ram-pressure stripping in galaxies, which our models suggest are not common. 

Despite these caveats, it is rather remarkable that reasonable agreement is found between different models and observational data sets. It is also surprising that models with as few as two free parameters can match galaxy colours and absorption linestrengths from cluster centre to outskirts, though simultaneously matching both observations is difficult even if pre-processing is neglected. Nonetheless, all of our successful models favour relatively gentle environmental quenching with timescales longer than the dynamical time in the cluster and longer than typically quoted for ram-pressure stripping of cold gas discs.

\section{Summary}
\label{sec:conclusions}

We have constructed a catalogue of subhalos in dark matter simulations of four rich cluster of galaxies. By tracking halo orbits, we have demonstrated that the infall history of halos imprints a relationship between two key parameters (infall time and halo velocity dispersion) of halos and their projected distance from the cluster centre. By assigning stellar masses to a single galaxy within each halo, we have also been able to test a variety of models for the star formation histories of cluster galaxies.  Novel aspects of this analysis are, first, the comparison with stellar absorption linestrengths and second, the separate consideration of bulge and disc colours, as opposed to global colours. Disc colours have been shown to be better discriminants of star formation models than total colours, because disc colours are most sensitive to environment.

In our best-fitting models, the bulge component of galaxies is quenched ``internally'', i.e.\ its age is not determined by infall into the cluster and can be fit with an old SSP. The observed colours of the discs of field galaxies require that, prior to infall, the star formation rate declines with a characteristic $e$-folding time $\tau\sbr{pre} = $ 5 -- 6.5 Gyr.

Models with no disc quenching fail to explain the redder colours of discs near the cluster centre. Models with short quenching times produce sharp changes in the median colours of discs and a stronger cluster-centric radial disc colour gradient than is observed, and so even ``delayed-then-rapid'' quenching models are disfavoured. Such models also overpredict the strength of the Balmer lines. Instead, we prefer a model in which the disc continues to experience exponentially declining star formation rates with shorter characteristic $\tau\sbr{post}=$ 3 -- 3.5 Gyr. This is suggestive of a gentler mechanism (e.g.\ ``strangulation'') as opposed to rapid ram-pressure stripping of the cold gas. Models where quenching only occurs within 0.5 $r_{200}$ produce better fits to the SDSS and NFPS disc colours.

In conclusion, we have found that although quenching is mild, our data require quenching to occur in cluster galaxies not long after they reach pericentre and on timescales of about 3 Gyr. Quenching may also take place in smaller clusters and groups, but it is not yet clear whether satellites must pass very close to a massive halo to be quenched, or whether infall anywhere within the virial radius is sufficient.

\section{Acknowledgments}
DST was supported in this work by the NSERC USRA program and Ontario Graduate Scholarships. MJH and MLB acknowledge support from their respective NSERC Discovery Grants. MLB acknowledges funding from NOVA and NWO visitor grants, which support his sabbatical visit to Leiden Observatory, where this work was completed. RJS was supported for this work by STFC Rolling Grant ST/I001573/1 ‘Extragalactic Astronomy and Cosmology at Durham 2008--2013’. The Centre for All-Sky Astrophysics is an Australian Research Council Centre of Excellence, funded by grant CE11E0090. Observational data used in this paper are available as raw imaging data from the NOAO data archives, and as parameters tabulated in the cited references.

\appendix
\section{Simple Models and the Effects of Mergers}
\label{sec:simplemergers}

\begin{figure*}
\includegraphics[width=0.50\textwidth]{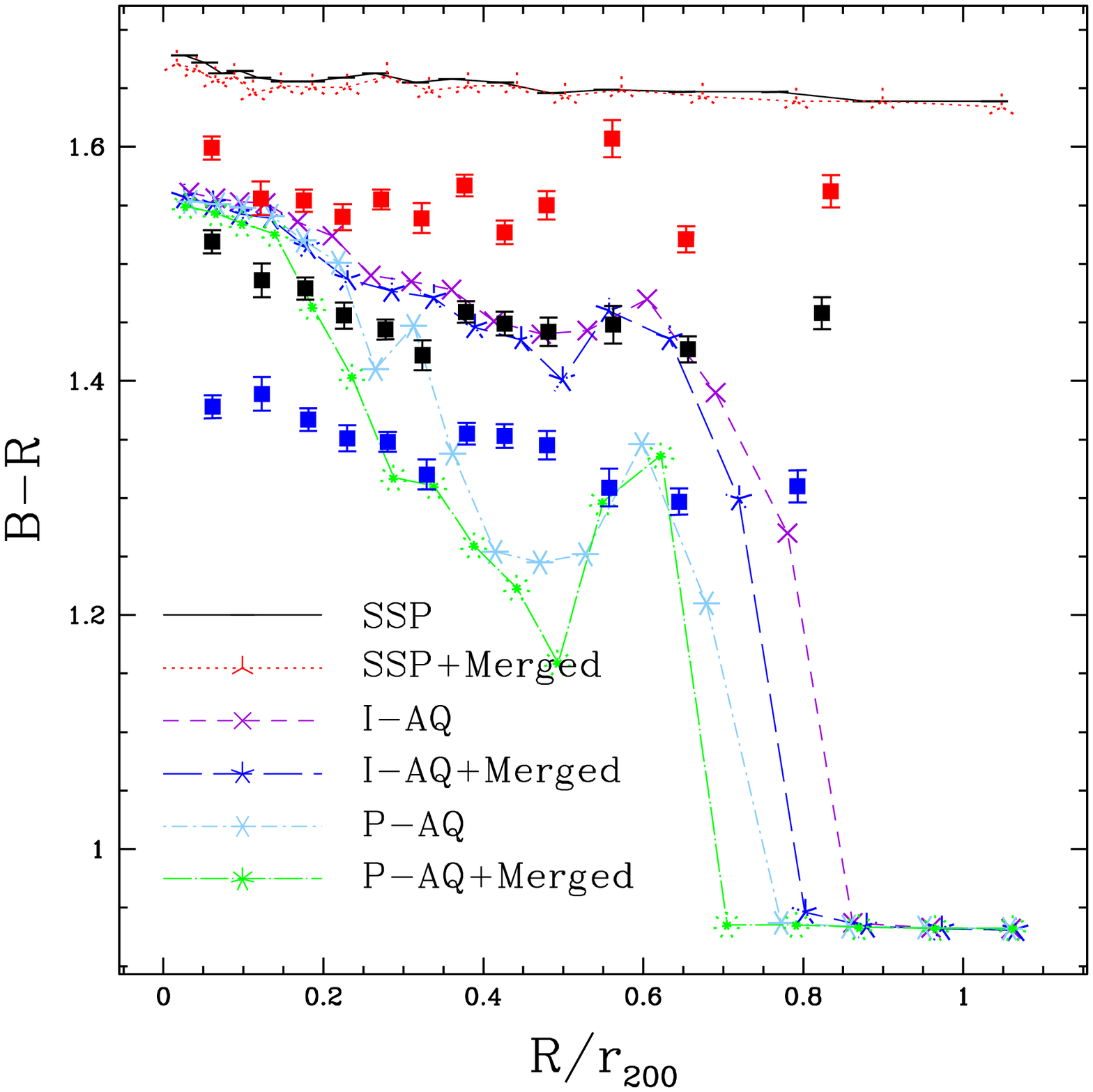}%
\includegraphics[width=0.50\textwidth]{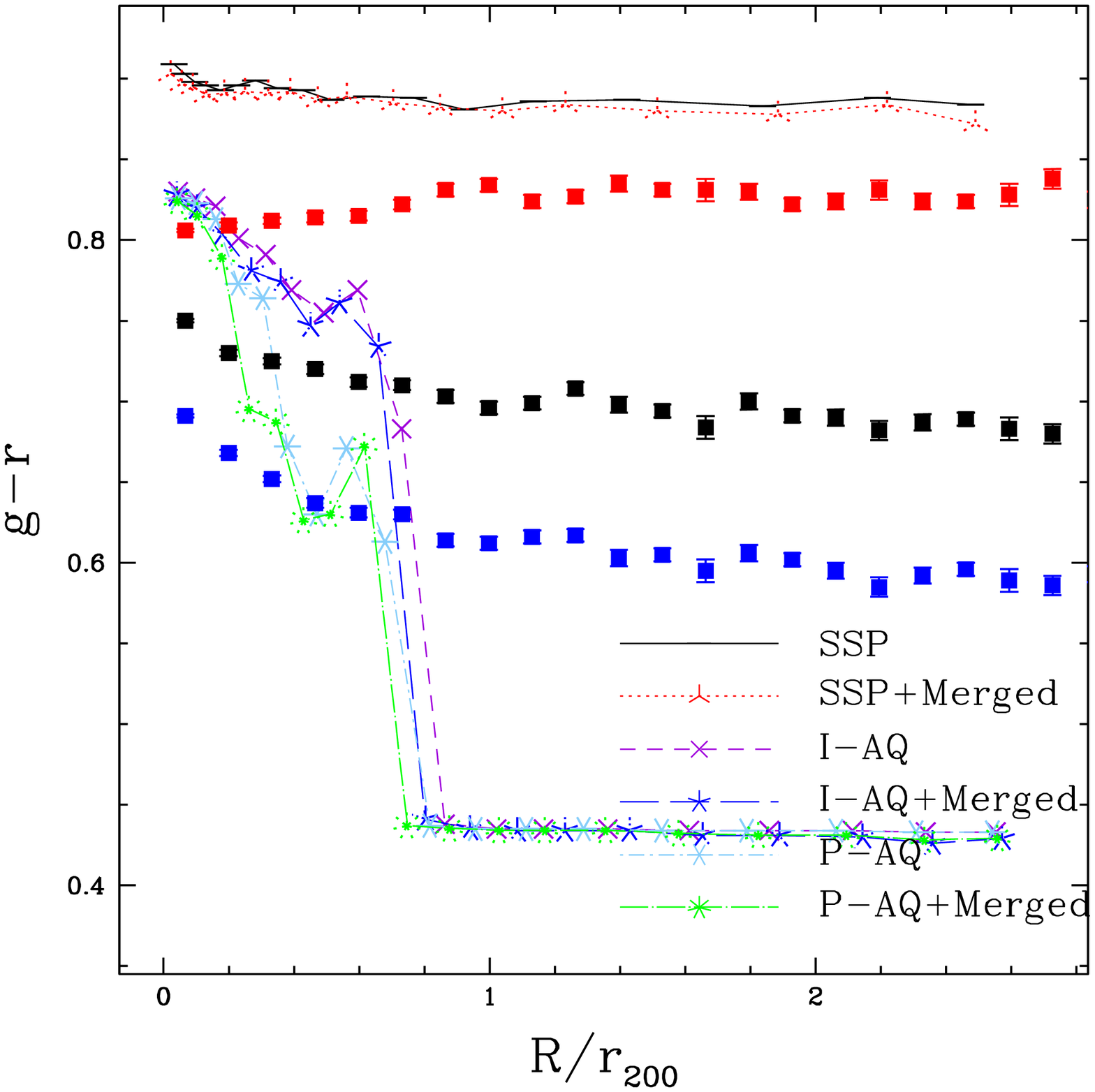}%
\caption{Simple, single-component star formation models described in \secref{singlemodels}, compared to NFPS $B-R$ colours (left) and SDSS $g-r$ (right). Observational data are median colours for bulge (red, uppermost values), disc (blue, lowest values) and total (black, middle values), with error bars representing errors on the medians in each bin. Three models are shown - an old, single stellar population (SSP), and two models with constant SFRs prior to infall followed by instant ``abrupt'' quenching (i.e. $\tau\sbr{pre}=\infty$, $\tau\sbr{post}=0$). The I-AQ model is quenched upon infall within $r_{200}$, while the P-AQ model is quenched at pericenter. Relations are also shown for models including likely mergers ('+Merged'). Candidate mergers are predominantly blue, so including these galaxies makes median colours at a fixed position slightly bluer, or equivalently, shifts the transition to blue-sequence dominated medians inwards. The effects are not large for infall-based models and negligible for mass-quenched models (bulges)
\label{fig:results_simple_merger}}
\end{figure*}

In \secref{simplemergers}, we presented basic methods for simple, single-component models based on either internal quenching or environmental quenching. We will now show the results of these models compared to observed trends in total, disc and bulge colours. 

Similarly, \secref{methods} presented a method for recovering overmerged orphan halos also recovered some genuine mergers. We identified candidate mergers using a somewhat arbitrary cut on distance (75 kpc) and binding ratio ($v/v\sbr{escape}$) to a more massive halo. We will now show the effects of this merger prescription on predicted colours for single-component models.

Colours for three single-component models are shown in \figref{results_simple_merger}. The three models are an internally-quenched SSP, and two ``abruptly-quenched'' (i.e. $\tau\sbr{post}$=0) models with constant SFR prior to quenching (i.e. $\tau\sbr{pre}=\infty$). The first model is quenched at infall past $r_{200}$ (I-AQ) while the second is quenched at pericenter (P-AQ). The internally-quenched model produces extremely red colours - even redder than bulge colours - and virtually no position-dependent trend, and so can only be considered a possible fit for bulge colours, not disc or total colours. The environmentally-quenched model can produce a wide range of trends in colours, but must be tuned to fit observed trends in disc colours. Pericenter quenching can alter the behaviour of the models at intermediate positions (0.2 $<R/r_{200}<$0.8) but does not alter the overall slope of the relation even in rather extreme model. We conclude that these models are overly simplistic, justifying the use of two-component models in \secref{twocomp}.

We now return to the merger prescription. Depending on the model, the fraction of bright, potentially merged galaxies is about 30-40\%. \figref{results_simple_merger} shows the results of including or potentially merged halos in single-component models (models including likely mergers are labelled '+Merged'). The differences are small, even for the most extreme abrupt quenching models, and excluding potential mergers generally makes median colours bluer, suggesting that candidate mergers are less likely to have been quenched than the remaining halos. 

The mass-quenched/bulge model is insensitive to the exclusion of candidate mergers. Two-component models are generally heavily weighted to the bulge component, so the effects on realistic bulge+disc systems are also much smaller than the offsets shown for a purely infall-quenched system. In summary, the effects of mergers are mainly to change the number of galaxies. Mergers have little impact on disc colours and a negligible impact on total colours.

\bibliographystyle{mn2e} 
\bibliography{paper_dt}

\end{document}